\documentclass[preprint]{aastex}
\usepackage{lscape} 
\usepackage{amsmath, amsthm,amssymb}
\usepackage{epstopdf}

\title{The Size Distribution of Kuiper belt objects for $D\gtrsim 10$ km}

\author{Wesley C. Fraser {1,2},
 JJ. Kavelaars {2,1}}
\altaffiltext{1}{Dept. of Physics and Astronomy, University of Victoria, Victoria, BC V8W 3P6, Canada}
\altaffiltext{2}{Herzberg Institute for Astrophysics, 5071 West Saanich Road Victoria, BC V9E 2E7, Canada}
\date{} 

\begin{abstract}
We have performed a  survey of the Kuiper belt covering $\sim 1/3$ a square degree of the sky using Suprime-cam on the Subaru telescope, to a limiting magnitude of $m_{50}(R)\sim 26.8$ and have found 36 new KBOs. We have confirmed that the luminosity function of the Kuiper belt must break as previously observed (Bernstein et al. 2004; Fuentes \& Holman 2008). From the luminosity function, we have inferred the underlying size distribution and find that it is consistent with a large object power-law slope $q_1\sim4.8$ that breaks to a slope $q_2\sim1.9$ at object diameter $D_b\sim60$ km assuming $6$\% albedos. We have found no conclusive evidence that the size distribution of KBOs with inclinations $i<5$ is different than that of those with $i>5$. We discuss implications of this measurement for early accretion in the outer solar system and Neptune migration scenarios.
\end{abstract}

\begin{document}

\maketitle

\newpage

\section{Introduction}
The Kuiper belt is a population of planetesimals with diameters as large as a few thousand kilometers \citep{Brown2006} and a mass of approximately a few $0.01 M_\oplus$ \citep{Gladman2001,Bernstein2004,Fuentes2008}. The state of the belt is enigmatic, as such large objects are not likely to form in such a low mass Kuiper belt, over the age of the solar system \citep{Kenyon2001}. Rather, it is likely that a much more massive initial belt underwent accretion for a time before some large-scale mass-depletion event occurred, such as a stellar passage \citep{Ida2000,Levison2004}, a sweep through of the mean-motion resonances of Neptune \citep{Levison2008}, or the scattering of KBOs by rogue planets \citep{Gladman2006}. Whatever the event, it is also responsible, at least in part, for the dynamically excited orbits of KBO populations such as the plutinos or the scattering population \citep{Gladman2008}.
 
The size distribution of the Kuiper belt contains a fossil record of the end-state of the accretion processes that occurred in that region. Knowledge of the size distribution can constrain disruption strengths of the bodies, formation time-scales in the outer solar system, and the early conditions of the proto-planetary disk. This makes the determination of the size distribution a primary constraint on Kuiper belt formation scenarios.

Because of the large distance to the Kuiper belt, the size distribution is not determined directly, but rather, is inferred from the shape of the observed luminosity function (LF). Early observations determined that the LF for bright objects $(m(R)\lesssim 26)$ was well represented by a power-law with a logarithmic slope $\alpha\sim 0.7$ \citep{Jewitt1998,Gladman1998,Gladman2001,Allen2002,Petit2006,Fraser2008}. This suggests that the size distribution of KBOs is a power-law with logarithmic slope $q\sim4.5$. The deepest survey of KBOs on the ecliptic, presented by \citet{Bernstein2004}, found a derth of faint objects demonstrating that the LF must ``roll-over'' or break to a shallower slope at $m(R)\lesssim27$ \citep{Bernstein2004}. Interpreting the roll-over in terms of the Kuiper belt size distribution implies that the large-object power-law size distribution breaks at a KBO diameter $D_b\sim 100$ km assuming 6\% albedos. Models of accretion in the outer solar system predicted such a break at least an order of magnitude smaller \citep{Kenyon2001,Kenyon2002}. 

The goal of our work was to confirm the results of \citet{Bernstein2004}, and accurately measure the shape of the LF at and beyond the roll-over magnitude. From this, the shape of the under-lying KBO size distribution can be inferred. We present here a survey of the ecliptic with limiting magnitude $m(R)\sim 27$, and use this along with previous observations to determine the shape of the LF. In section 2 we present our observations and in sections 3 we present our survey results. In sections 4 and 5 we consider past ecliptic Kuiper belt surveys and consider various LF functional forms. In sections 6 and 7 we present our analysis and a discussion of the results.

\section{Observations, image processing, and characterization}
\begin{figure}[h] 
   \centering
   \plotone{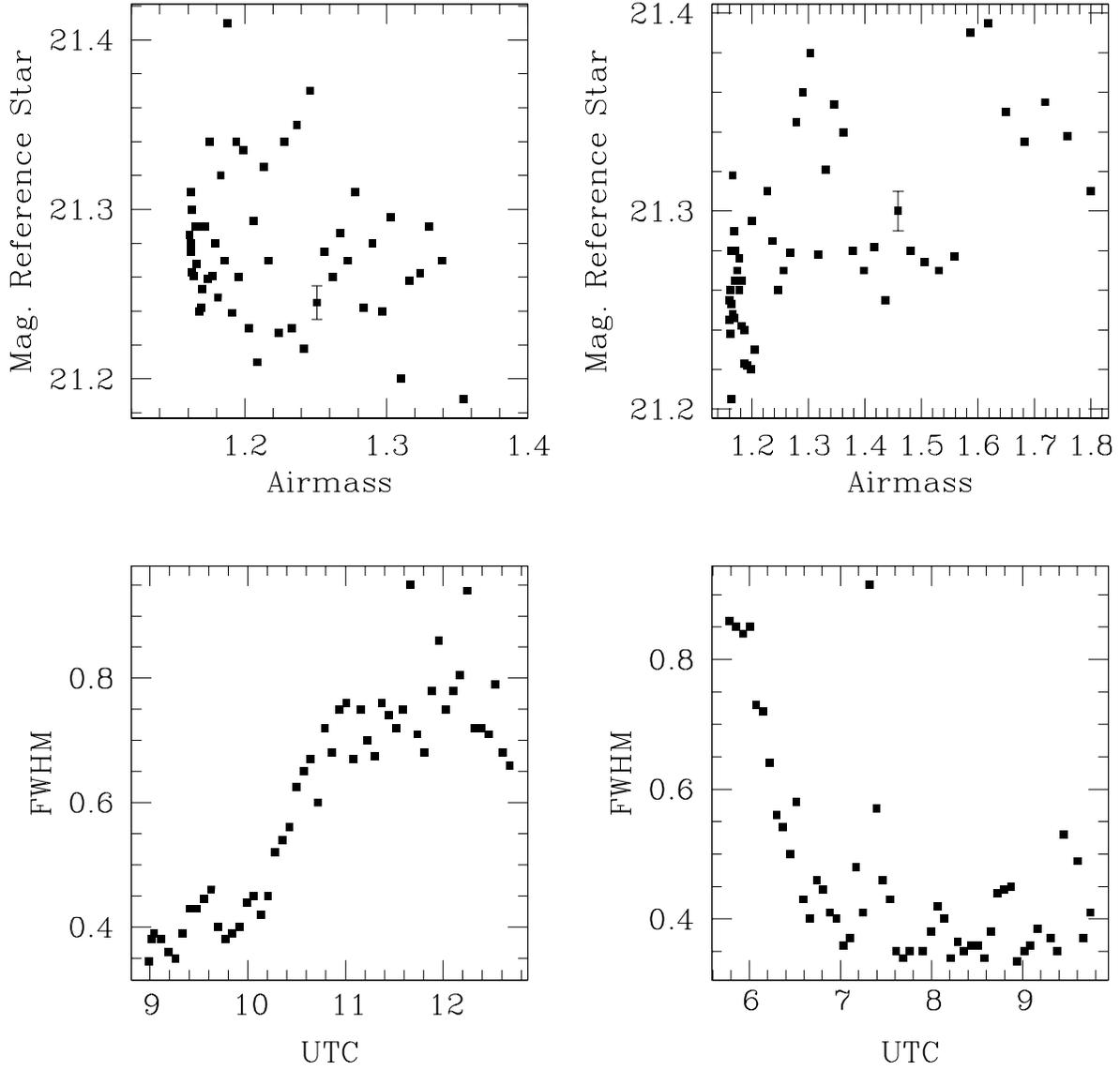} 
   \figcaption{Reference star magnitude versus airmass (top) and full-width-half-maximum versus observation time (bottom) for the April 22nd (left) and May 8th (right) nights. Typical uncertainty in the the magnitudes is shown. The reference star displays a $\sim0.1$ mag. variation in each night, which is smaller than the typical uncertainty in the flux measurements of each of our sources. \label{fig:seeing}}   
\end{figure}

New observations were made in service mode with Suprime-Cam on the Subaru 8.2 m telescope \citep{Miyazaki2002}. Suprime-cam is a 10 chip mosaic CCD camera with a 34'$\times$27' field-of-view (FOV) and a $0''.202$ pixel scale. Observations were made in the r-band on the nights of April 22nd (night 1), and May 8th (night 2) 2008. The same target field was observed each night, targeting the ecliptic with coordinates $\alpha=\mbox{13}^h\mbox{46}^m\mbox{57}^s\mbox{.7}$, $\delta=-\mbox{10}^o\mbox{44}^{'}\mbox{00}^{''}$. Additionally, each night, a photometric reference image was taken of the Canada-France-Hawaii Supernova Legacy Survey D2 field \citep{Astier2006}. Details of the observations are presented in Table ~\ref{tab:observations}. Presented in Figure~\ref{fig:seeing} is a summary of the image quality of both nights; as can be seen, the nights were photometric. 

Here we quickly describe our image processing techniques and subsequent moving object search for these data. A more thorough discussion of these techniques can be found in  our previous work \citep{Fraser2008}. We describe in more detail here the additional image processing steps that were necessitated by the use of Suprime-cam.

All images went through the same image pre-processing before the image search step. From the overscan strip, the bias level was measured for each image, and subtracted. It was found that the bias level varied by $\sim100$ ADU (10\% of the typical bias value) along the bias strip. Proper removal of this variation required that a fourth order single-dimensional polynomial be fit in a least-squares sense to the overscan region collapsed to a single column, and was done on an image-by-image basis. This polynomial bias-fit was removed from the raw images. 

An average flat-field was produced for each chip from the sky-flats provided from the service mode observer (10 from night 1, 6 from night 2) using the \textit{mask\_mkflat\_HA} routine from the sdfred software reductions package. This is the standard image pre-processing package provided and supported by Subaru \citep{Yagi2002,Ouchi2004}. This routine detects sources in the sky-flat images. Regions near detected sources are ignored when the average flat is produced. 

The Suprime-cam field is vignetted, with $\sim40\%$ throughput-loss near field corners. This vignetting needed to be accounted for when adding artificial sources to the data as the artificial sources must accurately reflect the sensitivity variations across each chip. Thus, the average flat-fields for each chip were separated into a quantum-efficiency variation map, and an illumination pattern. The illumination pattern was created by running a 20 by 20 pixel median filter over the average flat-fields. The quantum-efficiency variation map was set as the normalized difference of the flat-field and the illumination pattern, and was divided out of each of the over-scan subtracted images producing smooth images that revealed the vignetting pattern.

The Suprime-cam FOV is largely distorted such that a Kuiper belt object (KBO), with a rate of motion $\gtrsim 3$''/hr., whose motion would be linear on the sky, would not be linear in the images. Thus, spatial distortions needed to be removed. Image distortions were measured with the \textit{dofit} routine \citep{Gwyn2008}. This routine compares a reference source list containing true astrometric positions of sources, and positions of those sources within the images; from this, it calculates a 3rd order spatial distortion map. The image of the D2 field used to photometrically calibrate our images also provided a perfect astrometric reference, as sources in the field have been astrometrically calibrated with residuals better than 0.3''. The distortion map was measured from the D2 field image of each night, and then used to resample all images onto a spatially flat image using the \textit{SWarp} package \citep{Bertin2002}. Flux was conserved during this resampling, and produced images with a 0.1986" pixel-scale. The spatial rms residuals of the resampling were 0.06", or $\sim 1/3$ of a pixel as confirmed using the astrometric positions of stars in the USNO catalogs.

The point-source image shape varied significantly across each chip, and warranted the use of a spatially variable point-spread-function (PSF). It was found that a PSF whose shape varied linearly with image position could accurately re-produce the point-source image shape variations. For each chip, a gaussian PSF with look-up table was generated from 15 bright, unsaturated, hand-selected point-sources. 

For each chip, a random number of artificial moving sources (between 150 and 250) were generated with random rates and angles of motion consistent with objects on circular Keplerian orbits between 25 and 200 AU, and with random fluxes consistent with that of point sources between 23 and 28.5 mag. These artificial sources were implanted blindly in the data, and the source list was revealed only after the search was complete. Additionally, 10 23rd mag. sources were implanted. These sources had flux sufficient to flag errors in the image combining algorithms.

Artificial sources with sky motions larger than 0.2 pixels were split up into a number of dimmer sources with total flux equal to the original source. The centres of the dimmer sources were shifted to account for the motion of the object in the images. The flux of each artificial source was varied from image to image to match the average brightness variations of 20 reference stars with respect to a reference image (that with the lowest airmass). Additionally, the fluxes of the artificial sources were scaled to account for the vignetting apparent in each chip. This was done by implanting the artificial source into a blank image. The blank image pixel values were then multiplied by the illumination pattern. This was subsequently summed with the vignetted sky images. After artificial object implanting, the illumination pattern was removed from the images. The result was images with spatially smooth backgrounds containing artificial sources with spatially varying sensitivity that matched that of the images.

No attempt was made to subtract a bias pattern from the raw images as no bias frames were provided from the service observations. The sdfred documentation however, claims that this step is unnecessary, as the bias level variations of Suprime-cam are low \citep{Yagi2002,Ouchi2004}.

To maximize the searchable area of the images, stationary objects were removed by subtracting an image template from the  vignetting corrected images. The image template for each chip was created from an average of the search field images using an artificial skepticism routine which places little weight on pixels far away from the pixel average \citep{Stetson1989}. Every fifth image was used to create the image template, reducing the presence of subtraction residuals created behind all moving sources, while ensuring a high quality template such that all stationary sources were sufficiently removed. The image subtraction was done using \textit{psfmatch3} routine \citep{Pritchet2005}. 

Before final image stacks were produced, a high pixel mask was applied to the subtracted images to mask out any spurious hot-pixels and cosmic-ray spikes. The masked-subtracted images were spatially shifted to account for the motions of moving sources.  A grid of  shift rates ($0.4-4.5$ "/hr.) and angles ($\pm 15^o$ off the ecliptic) were considered which covered the range of motions consistent with bound solar system objects in the Kuiper belt on prograde orbits. In our past searches, we found that the search depth is not very sensitive to the choice in grid spacing between angles, but is very sensitive to the choice in spacing between rates. We chose a rate spacing small enough such that a moving source would never exhibit a trail longer than twice its seeing disk at the rate and angle which best approximated its apparent motion. Our grid consisted of 19 separate rates and 5 separate angles.

The data from each night were photometrically calibrated to the SDSS wavebands by comparing source flux measurements in the D2 field to those quoted in the Mega-pipe project \citep{Gwyn2008}. A source's magnitude in the Subaru data is given by $r_{Sub} = -2.5 \log \left(\frac{b}{t}\right) +Z$ where $b$ is the source brightness in ADU, $t$ is the exposure time, and $Z$ is the telescope zeropoint. For each chip, a linear relation of the form 
$r_{Sub} = C(g'-r')_{Mega} - r'_{Mega}$ 
was used to transformation the r'-band magnitudes in the Mega-pipe project to Subaru r-band magnitudes. The colour term, $C$, and the zeropoint $Z$ were fit in a least-squares sense to the source-flux measurements in the data, and the Mega-pipe magnitudes using at least 40 sources on each chip. This procedure determined the best fit zeropoint to an accuracy of $\sim0.04$ mag. The results are presented in Table~\ref{tab:calibrations}. We found that chip 00 exhibits a $~0.4$ magnitude decrease in sensitivity compared to the other 9 chips, which is consistent with other photometrically calibrated Suprime-Cam observations \citep{Yoshida2007}.

The transformations of the r' and g'-band Mega-pipe magnitudes to the r', and g'-bands in the SDSS filter set are given by $r'_{Mega}=r'_{SDSS} - 0.024 (g'-r')_{SDSS}$ and $g'_{Mega}=g'_{SDSS}-0.153 (g'-r')_{SDSS}$. Using the colour term $C_{av}=-0.048$ which is the average of colour terms from Table~\ref{tab:calibrations}, we find that the transformation between r-band Subaru and SDSS magnitudes is given by $r'_{SDSS} = r_{Sub} +0.018 (g'-r')_{SDSS}$. For typical KBOs $(g'-r')_{SDSS} \sim 0.7$ mag \citep{Fraser2008}. Thus, the r-band Subaru and SDSS magnitudes differ by $\sim 0.01$ mag. which is smaller than the uncertainty of the telescope zeropoint. Hence, we use the approximation that for KBOs, $r'_{SDSS}=r_{Sub}$.

\section{Survey and results}
From the first night, the grid of image stacks for each chip was manually searched by one operator producing a candidate source list. Moving sources were identified by their appearance in the images; a moving source is round (or nearly) in the image stack that best compensated for the source's apparent sky motion, and exhibits a characteristic trail in other stacks, helping the operator distinguish moving sources from noise. See \citet{Fraser2008} for a more complete discussion of this technique.

\begin{figure}[t] 
   \centering
   \plotone{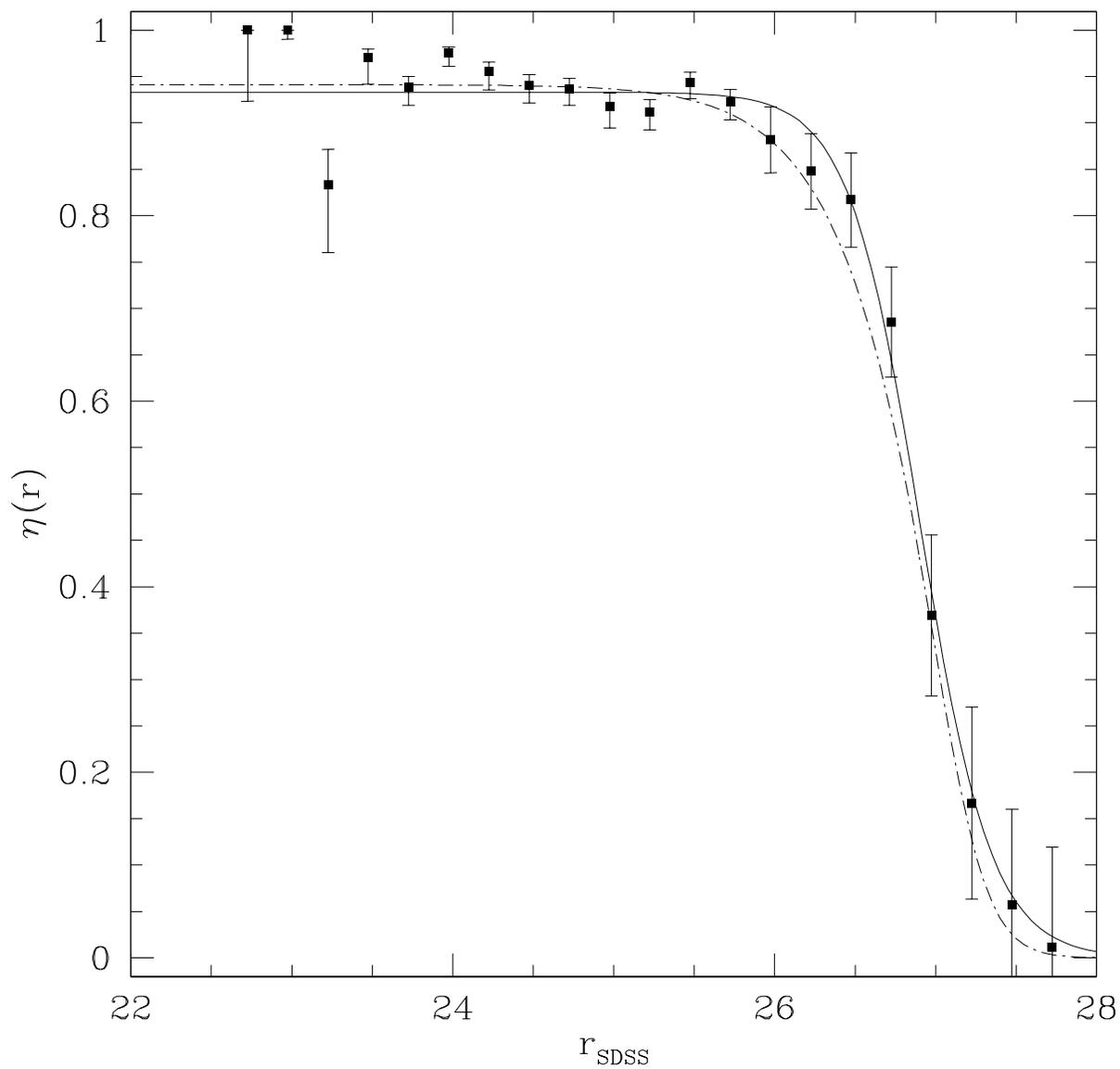} 
   \figcaption{Detection efficiency histogram with 1-sigma poisson errorbars, and the best-fit efficiency functions $(\eta_{max},r_*,g)= (0.932, 26.86, 0.44)$ (solid) and $(\eta_{max},r_*,g,g')=(0.941,27.14,0.333,0.758)$ (dashed-dotted). \label{fig:eff}}   
\end{figure}

Implanted artificial sources were identified in the candidate source list if the candidate source's location was within a few pixels of the artificial source's true centre. The detected artificial sources allowed us to measure our detection efficiency as a function of magnitude, $\eta(r'_{SDSS})$. We represent the detection efficiency of the full field by the functional form 

\begin{equation}
\eta(r'_{SDSS})=\frac{\eta_{max}}{2}\left(1-\tanh\frac{r'_{SDSS}-r_{*}}{g}\right)
\label{eq:eff}
\end{equation}

\noindent
where $\eta_{max}$ is the maximum efficiency, $r_*$ is the half-maximum detection efficiency magnitude, and $g$ is approximately half the width of the fall from maximum detection efficiency to zero. This function was fit, in a least-squares sense, to the detection efficiency of artificial sources. The best fit parameters were $(\eta_{max},r_{*},g) = (0.932\pm0.008, 26.86\pm0.02, 0.44\pm0.03)$. The best-fit curve represents the measured detection efficiency well (see Figure~\ref{fig:eff}). 

We fit the measured efficiency with $\eta'(r'_{SDSS})=\frac{\eta_{max}}{2}\left(1-\tanh\frac{r'_{SDSS}-r_{*}}{g}\right)\left(1-\tanh\frac{r'_{SDSS}-r_{*}}{g'}\right)$ which has been found to represent the detection efficiency of previous KBO surveys \citep{Petit2006,Fuentes2008}. The best-fit curve is presented in Figure~\ref{fig:eff}. As can be see, there is very little difference between the more complicated efficiency representation and that of Equation~\ref{eq:eff}, and both describe the measurements equally well. Thus, we find that the more complicated function is not warranted or necessary.

Flux measurements were made for all identified moving sources on the shifted image stack which contained the roundest image for that source. Magnitudes were measured in a 3-4.5 pixel radius aperture, and corresponding aperture corrections were determined from the image profiles of ten 23rd mag. artificial sources on each chip. By comparing the flux measurements of the detected artificial sources to their true values, we were able to characterize our flux measurement uncertainties. For background limited sources, the uncertainty in a source's measured magnitude, $r$, is given by $\Delta r =\gamma 10^\frac{r-Z}{2.5}$ where $Z$ is the telescope zeropoint, and $\gamma$ depends on the telescope and the observations \citep{Fraser2008}. We found that $\gamma = 1.33$ best described the observed uncertainties.

38 sources were not associated with artificial sources. 36 of these are identified as newly discovered KBOs. Two of these were identified as coincidental overlap of multiple poorly subtracted saturated galaxies. Both of these false detections have fluxes below the 50\% detection efficiency. This is consistent with our previous survey - a non-zero false candidate rate occurred for detection efficiencies $\eta<50\%$ \citep{Fraser2008}. We therefore truncate our detection efficiency at the 50\% threshold and ignore the 8 sources faint-ward of this level.

From the positions of the objects on the first night, orbits were calculated using \textit{fit\_radec} \citep{Bernstein2000} and positions of the objects on night 2 were predicted. While not all of our objects had follow-up detections, we had a 100\% follow-up rate for those 7 objects whose predicted positions fell in the FOV of the second night's observations. This is consistent with the 100\% follow-up rate above the 50\% detection threshold we report from our past survey \citep{Fraser2008}. We are thus confident that all detected non-artificial sources with magnitudes above the 50\% threshold are real KBOs.

For each detected KBO, three image stacks were created (6 for those with follow-up) of the first half, middle half, and last half of the vignetting removed, non-subtracted images from a night's observations, using the rate and angle of motion that produced the roundest source image. From these, flux measurements were made. The final source magnitudes we report are the average of all the flux measurements made for that source off the image stacks where the source was at least a few FWHM away from nearby bright stars and galaxies. The results of our survey are presented in Table~\ref{tab:objects}.

\section{Luminosity function}
Previous surveys have found that the differential LF (number of KBOs with magnitude $m$ to $m+dm$ per square degree) of bright KBOs ($m(R)\lesssim 26$) mag.  is well represented by 

\begin{equation}
\Sigma(m)=\ln(10) \alpha 10^{\alpha(m-m_o)}
\label{eq:powerlaw}
\end{equation}

\noindent
where $\Sigma(m)$ is the number of objects with magnitude $m$ (usually R-band) per square degree, $\alpha$ is the power-law ``slope'', and $m_o$ is the magnitude for which the sky density of objects with magnitudes $\leq m$ is one per square degree (see for example \citet{Trujillo2001b,Gladman2001,Petit2006}). In our previous work, we found  $\alpha=0.65$ and $m_o(R)=23.43$ for a broad range of surveys \citep{Fraser2008}. \citet{Bernstein2004} presented the faintest Kuiper belt survey to date. Using a substantial amount of Hubble observations ($\sim 100$ orbits) they achieved a search depth of $m(R)\sim 28.5$ \citep{Bernstein2005} and discovered three KBOs. The dearth of detections was at least a factor of $\sim10$ below the number expected, requiring that the LF has a steep logarithmic slope for bright objects, that ``rolls-over'' to shallower slopes for fainter objects.

\citet{Fraser2008} presented a survey in which they searched 3 square degrees of the ecliptic to a depth of $m(R)\sim25.4$ mag. and substantially increased the number of known detections suitable for a determination of the LF faint-ward of $m(R)\sim 23.5$. They found that the LF was well represented by a power-law given by Equation~\ref{eq:powerlaw}. They concluded that the lack of objects observed by \citet{Bernstein2004} must be caused by a sharp break rather than a broad roll-over, and that this could not occur for $m(R)\leq 24.4$. 

Additionally, \citet{Bernstein2004} examined the LFs of the so-called cold and excited KBO populations, $i < 5^o$ and $i>5^o$ respectively. They found that the cold population exhibited a steeper bright object slope than the excited population.  The statistical support for such a difference however, was not strong. Simulations that test different Kuiper belt formation scenarios suggest that a difference  in the SDs of different KBO populations is caused by different formation conditions/histories for those objects. Thus, a confirmation of the findings by \citet{Bernstein2004} is important, as it provides a very strong constraint on the Kuiper belt formation history.

The survey reported here detected 28 KBOs that can be used to characterize the luminosity function (LF) and SD of KBOs. With these detections, we wish to address three questions:

\begin{enumerate}
\item At what magnitude is the large object LF no longer a power-law?
\item Is the observed break consistent with the LF expected if the SD has a sharp break at some object radius? If not, what is the true shape of the LF?
\item Is the SD of the excited population different from that of the cold population?
\end{enumerate}

\subsection{Luminosity Function forms}
In determining the correct LF shape, we consider three functional representations of the differential LF. The first is the power-law given by Equation~\ref{eq:powerlaw}. The second is the rolling power-law suggested by \citet{Bernstein2004}, given by

\begin{equation}
\Sigma(m) = \Sigma_{23} 10^{\alpha(m-23)+\alpha'(m-23)^2}
\label{eq:rolling}
\end{equation}

\noindent
where $\alpha$ is the bright object slope, $\alpha'$ is the derivative of the logarithmic slope, and $\Sigma_{23}$ is the number of objects per square degree at 23rd mag.

The third functional form we consider was first presented by \citet{Fraser2008b}. They showed that, if the SD is a power-law with slope $q_1$ for large objects that has a sharp break at objects with diameter $D_b$ to slope $q_2$ for small objects, then the \textit{cumulative} LF has the form

\begin{equation}
N\left(<m\right) = A
\begin{cases}
 a_1 10^{\alpha_1 m} & \mbox{if $m < m_b$}\\
 b_110^{\alpha_2 m} + b_2 10^{\alpha_1 m} + b_3\left(m\right) & \mbox{ if $m_b \leq m \leq m_2$}\\
 c_1  +c_2 10^{\alpha_2 m} & \mbox{ if $m > m_2$}.
 \end{cases}
 \label{eq:Fraser2008}
 \end{equation}

\noindent
where $\Sigma(m) = \frac{dN(<m)}{dm}$, $\alpha_1=\frac{q_1-1}{5}$ and $\alpha_2=\frac{q_2-1}{5}$. In Equation~\ref{eq:Fraser2008} $m_b=K+5\log \left(r_1^2 D_{b}^{-1}\right)$ (break magnitude) and $m_2=K+5\log \left(r_2^2 D_{b}^{-1}\right)$ are the magnitudes for which objects smaller than the break diameter are detectable by a given survey for heliocentric distances larger than the Kuiper belt inner edge $r_1$ but smaller than the outer edge $r_2$. 

The parameter $K$ is a constant relating the diameter of an object to its apparent magnitude. Choosing a value of $K$ is equivalent to choosing a constant albedo for KBOs. It is apparent that the largest objects have albedos as high as $\sim90$ \%, with smaller objects exhibiting much lower values \citep{Stansberry2007}. Clearly the assumption that albedos are constant is incorrect, at least when considering the largest KBOs. Insufficient knowledge of the albedos for smaller $D\lesssim 500$ km KBOs - the size range which we probe in this work - exists for this assumption to be tested. \citet{Fraser2008} discuss the effects of a varying albedo when a constant is assumed, and demonstrate that the inferred size distribution slope is incorrect. This effect however, is small compared to the uncertainties typical in the LF determination, given the available data.

It is apparent, at least for the observed sources, that an albedo of 6\% is a typical value for $D\lesssim 500$ km KBOs \citep{Stansberry2007}, and is the value we assume here. This sets $K\sim18.4$ as the R-band magnitude of a $D\sim1600$ km object at 40 AU with a 6\% albedo. This apparently typical albedo might very well be an observational bias. Until more data are available, it will not be known whether 6\%, or a much larger value similar to large objects, is a more appropriate choice. Thus, any conclusions about the inferred break size come with this caveat; the real break diameter could be quite different dependening on the true albedos.

This functional form has the advantage that the LF can be interpreted in terms of LF parameters $(\alpha_1,\alpha_2,m_b,\log A)$ or SD parameters $(q_1,q_2,D_b,\log A)$. Both provide equivalent representations of the LF shape, and, assuming an albedo, the SD is directly inferred from the best-fit LF.

The coefficients $a_i$, $b_i$, and $c_i$ of Equation~\ref{eq:Fraser2008} are functions of the KBO radial distribution. \citet{Fraser2008b} have shown that the LF shape for $m_b\leq m \leq m_2$ is strongly dependent on the assumed KBO radial distribution, but the SD slopes $q_1$ and $q_2$, and the break diameter $D_b$ however, can be accurately inferred if data exists sufficiently far away from the break magnitude. They assume that the radial distribution is given by $N(\Delta) \propto \Delta^{-c}$, and suggest that $c=10$ is a good approximation to the radial fall-off observed by \citet{Trujillo2001}. We assume the same radial distribution here.

These three functions provide a good sample of LF forms to consider as they progress in levels of complication by a single degree of freedom - the parameters for the different LF forms are $(\alpha,m_o)$ for Equation~\ref{eq:powerlaw}, $(\alpha,\alpha',\Sigma_{23})$ for Equation~\ref{eq:rolling}, and $(\alpha_1,\alpha_2,m_b,A)$ for Equation~\ref{eq:Fraser2008}. Thus,  it is a straight-forward process to determine the number of parameters necessary to describe the true shape of the LF using simple statistics techniques.

\subsection{Data-sets}
In our previous estimates of the LF, we considered observations from surveys for which the detection efficiency was well characterized as a function of magnitude. The F08 sample \citet{Fraser2008} consists of near ecliptic KBO surveys presented in \citet{Jewitt1998,Gladman1998,Gladman2001,Allen2002,Petit2006}, and \citet{Fraser2008} along with the data from \citet{Trujillo2001b} which we subdivided by detection efficiency and astrometric position. The F08b sample is the F08 sample, plus the inclusion of the survey presented by \citet{Bernstein2004,Bernstein2005} and {\it the survey presented here}. 

To measure any differences in the LF between the dynamically excited and cold populations, we define the F08b$_{i<5}$ and F08b$_{i>5}$ samples as those objects with inclinations smaller than and greater than $5^o$ respectively. For these samples, we only consider surveys in which the majority of objects have been observed on arcs longer than 24 hrs, inclinations measured from the observations spanning only one night are not accurate enough to make an inclination cut reliable, and a large contamination between the two samples would occur. Thus we exclude from the F08b$_{i<5}$ and F08b$_{i>5}$ subsamples the Uranus field from \citet{Petit2006}, the N11033, N10032W3, UNE, and UNW fields from \citet{Fraser2008}, and the survey presented here.

While an inclination cut does not directly probe the different KBO populations, ie. cold/hot-classical objects, resonators, scattered members, etc. \citep{Gladman2008}, it does provide a crude means of separating those KBOs with `highly' excited orbits from those with less excited orbits. The division of $i=5^o$ is chosen to ease comparison with the LF presented by \citet{Bernstein2004}. For the F08b$_{i<5}$ and F08b$_{i>5}$ samples, no  heliocentric distance cut was made.

Note: the analysis for this manuscript was already complete before the authors were aware of the survey presented by \citet{Fuentes2008}. The moderate increase in the total number of detections, $\sim 20\%$, will not substantially alter the results presented here. Thus, we did not include the survey of \citet{Fuentes2008} in our analysis. Their results however, are discussed below.


\subsection{Analysis \label{sec:analysis}}
In determining the correct LF shape and best-fit parameters of the various functional forms, we adopt the same practices as \citet{Fraser2008,Fraser2008b}. These are as follows:

\begin{enumerate}
\item We cull from each survey all sources faintward of the 50\% detection efficiency of that survey, and set the detection efficiency to zero faintward of that point.
\item We offset all magnitudes to R-band using the average KBO colours presented in \citet{Fraser2008}.
\item We fit the differential LF, $\Sigma(m)=\frac{dN(<m)}{dm}$ using a maximum likelihood technique.
\item We adopted the same form of the likelihood equation as presented in \citet{Fraser2008} which treats calibration errors and variable sky densities as separate nuisance parameters. We marginalized the likelihood over these parameters before determining the maximum likelihood.
\item We adopted the same prior ranges of the colour and density parameters as presented in \citet{Fraser2008} ($\pm 0.2$ magnitudes for the colour parameters, and $\pm 0.4$ for the logarithm of the density parameters). 
\end{enumerate}
 
In determining the quality of the fits, we utilize the Anderson-Darling statistic

\begin{equation}
\Delta=\int_0^1 \frac{\left(S(m)-P(m)\right)^2}{P(m)\left(1-P(m)\right)}dP(m)
\label{eq:AD}
\end{equation}
\noindent
where $P(m)$ is the cumulative probability of detecting an object with magnitude $\leq m$ and $S(m)$ is the cumulative distribution of detections. We calculate the probability, $P(\Delta>\Delta_{obs})$ of finding a value of the Anderson-Darling statistic, $\Delta$, larger than that of the observations for a given LF function and parameter set by bootstrapping the statistic, is. randomly drawing a subsample of objects from that LF equal to that detected in a particular sample, and fitting the LF to that random sample and computing $\Delta$. In the random sampling process, we include random sky-density and colour offsets which represent those expected from the real observations. Values of $P(\Delta>\Delta_{obs})$ near 0 indicate that the functional form is a poor representation of the data. 

To determine if the more complicated LF functional forms are statistically warranted, we utilize the log-likelihood test $\chi^2=-2\log\frac{L'}{L}$. The logarithm of the ratio of maximum likelihoods of the simple and more complicated functions, $L$ and $L'$, are distributed as a chi-squared variable with a number of degrees of freedom equal to the difference in free parameters between the two functional forms. A table of $\chi^2$ values can be used to determine the significance of the improvement in the LF fit between the two functional forms.  \citep{Kotz1983}.

\section{Results \label{sec:Results}}
\subsection{F08b Sample}
Presented in Table~\ref{tab:fits} are the best-fit parameters for the three different LF forms, and $P(\Delta>\Delta_{obs})$ of each fit. The best-fit power-law slope to the F08b data-set, $\alpha=0.58$, is moderately shallower than, but is consistent with the best-fit $\alpha=0.65$ from \citet{Fraser2008}. The fit however, is a poor description of the data as evidenced by the low  $P(\Delta>\Delta_{obs})<0.02$. 

The best-fit of the rolling power-law to the F08b data-set is $(\alpha,\alpha',\Sigma_{23})=(0.8,-0.06,0.82)$. The maximum likelihood value of this fit is increased by more than two orders of magnitude over the power-law which warrants the inclusion of a third degree of freedom, and the fit is an acceptable description of the F08b sample, with $P(\Delta>\Delta_{obs})= 0.4$. 
 
The best-fit of the LF given by Equation~\ref{eq:Fraser2008}, $(\log A,\alpha_1,\alpha_2,m_b) = (23.56,0.76,0.18,24.9)$, is an acceptable fit to the data with $P(\Delta>\Delta_{obs}) = 0.4$. The maximum likelihood value is increased by more than two orders of magnitude over the power-law with a log-likelihood chi-square, $\chi^2 = 11.7$, and by a factor of 2 over the rolling power-law with $\chi^2=1.4$. We find that the best-fit from Equation~\ref{eq:Fraser2008} is preferred over the power-law at greater than the 3-sigma level. This fit is preferred over the rolling power-law at $\sim 80\%$ significance. Thus we find that both the rolling power-law and broken power-law of Equation~\ref{eq:Fraser2008} provide equally adequate descriptions of the F08b sample. Note: the Monte-Carlo simulations we have done to calculate $P(\Delta>\Delta_{obs})$ have also demonstrated that on average, the best-fit LF parameters determined from our maximum likelihood technique reproduce the input parameters of the simulations for all LF functionals forms we have considered here.

When the observations of \citet{Bernstein2004} are excluded from the F08b sample, the best-fit power-law has parameters $(\alpha,m_o)=(0.65,23.42)$, which is nearly identical to that found from by \citet{Fraser2008}. The power-law is a moderately adequate description of the observations with $P(\Delta>\Delta_{obs}) =0.1$. This result hints that the Kuiper belt LF exhibits a break within the magnitude range of the F08b sample, but fits to the survey data with $m(R)\lesssim27$ do not explicitly require a broken or rolling power-law description. 

\begin{figure}[t] 
   \centering
   \plotone{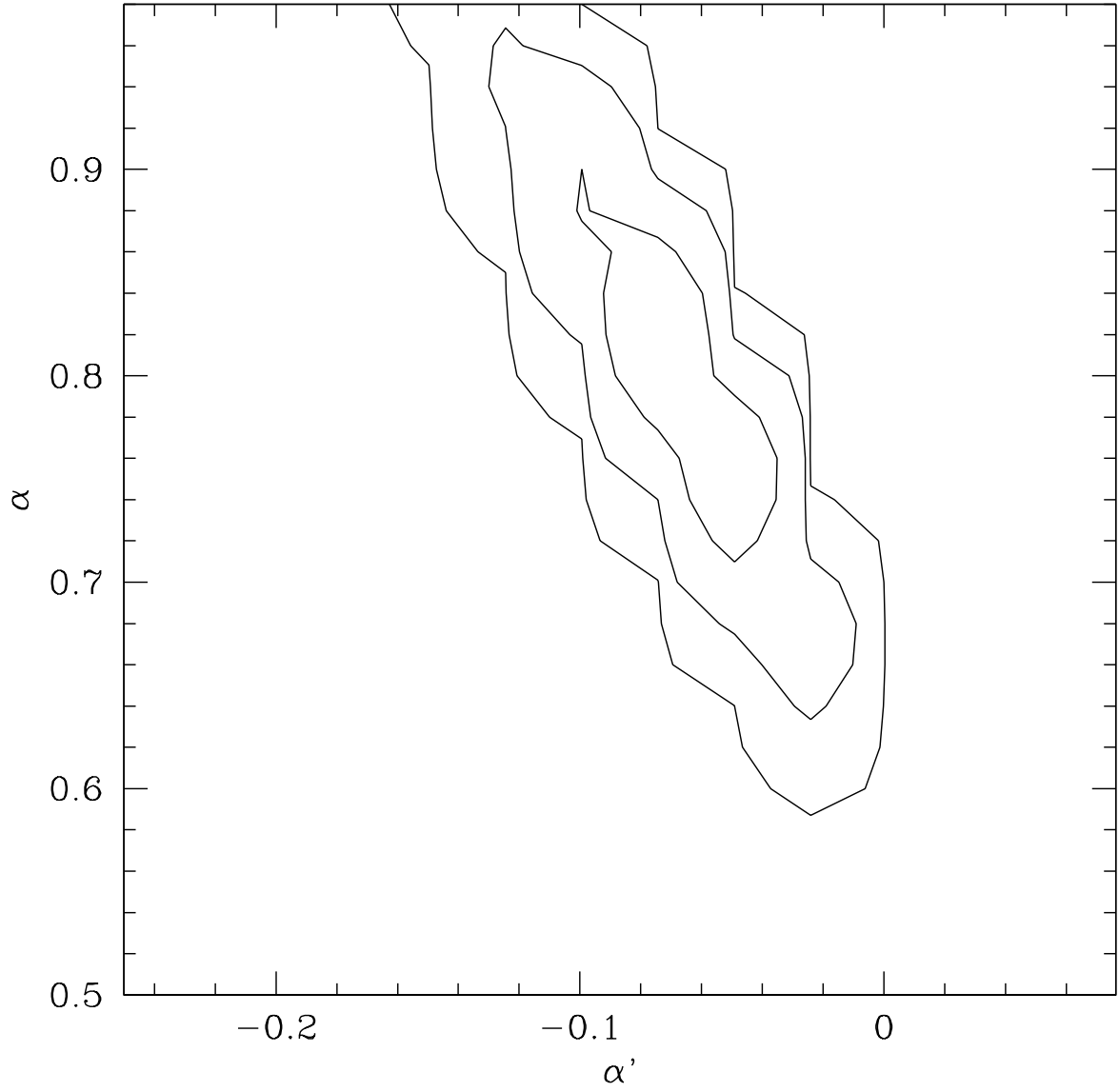} 
   \figcaption{The 1,2, and 3-sigma credible regions of the best-fit rolling power-law to the F08b sample with $\Sigma_{23}$ held at its best-fit value. \label{fig:a1ap}}   
\end{figure}

\begin{figure}[t] 
   \centering
   \plotone{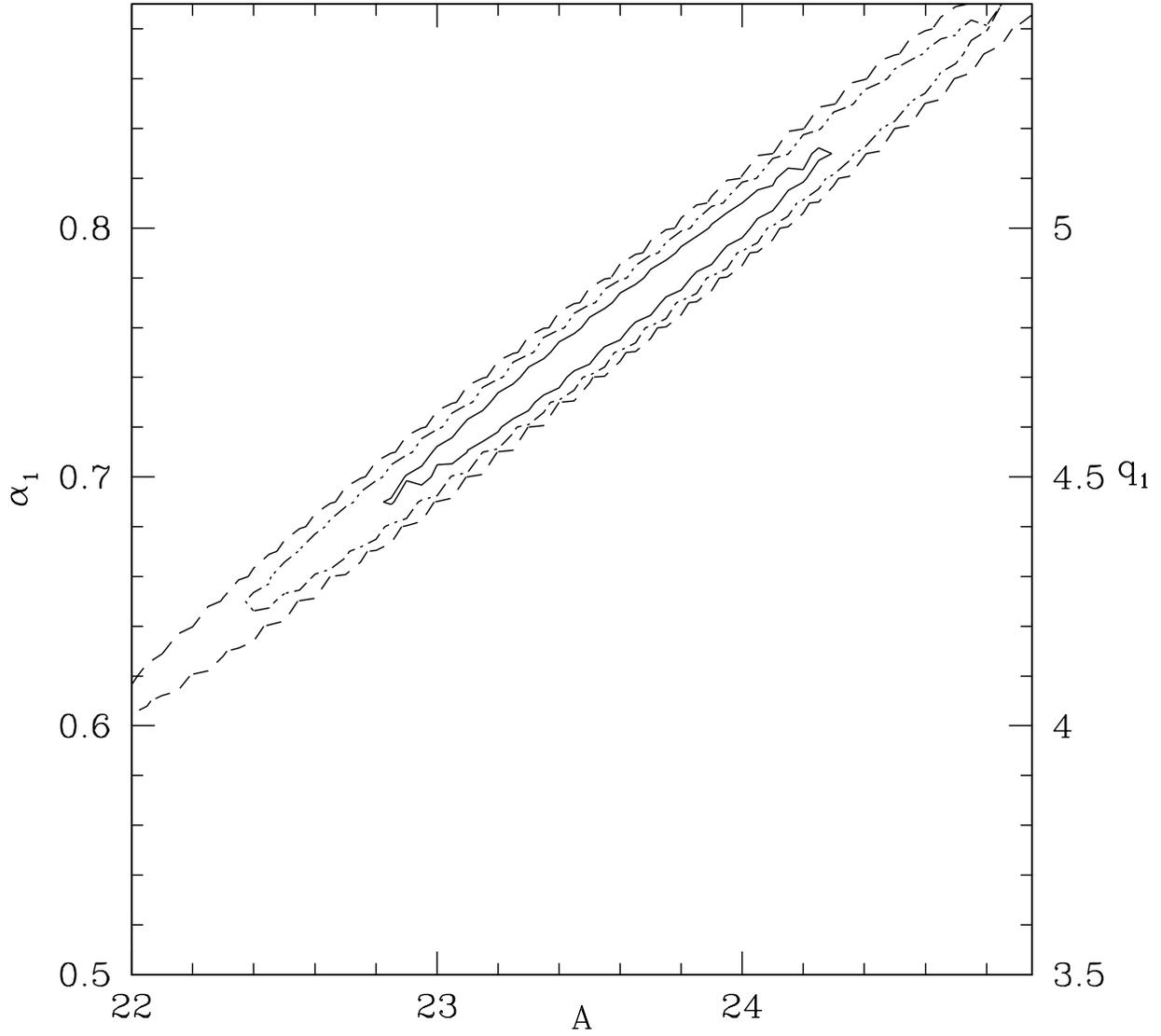} 
   \figcaption{The 1,2, and 3-sigma credible regions of the best-fit of Eq~\ref{eq:Fraser2008} to the F08b sample with $\alpha_2$ and $D_b$ held at their best-fit values. \label{fig:a1N}}   
\end{figure}

\begin{figure}[t] 
   \centering
   \plotone{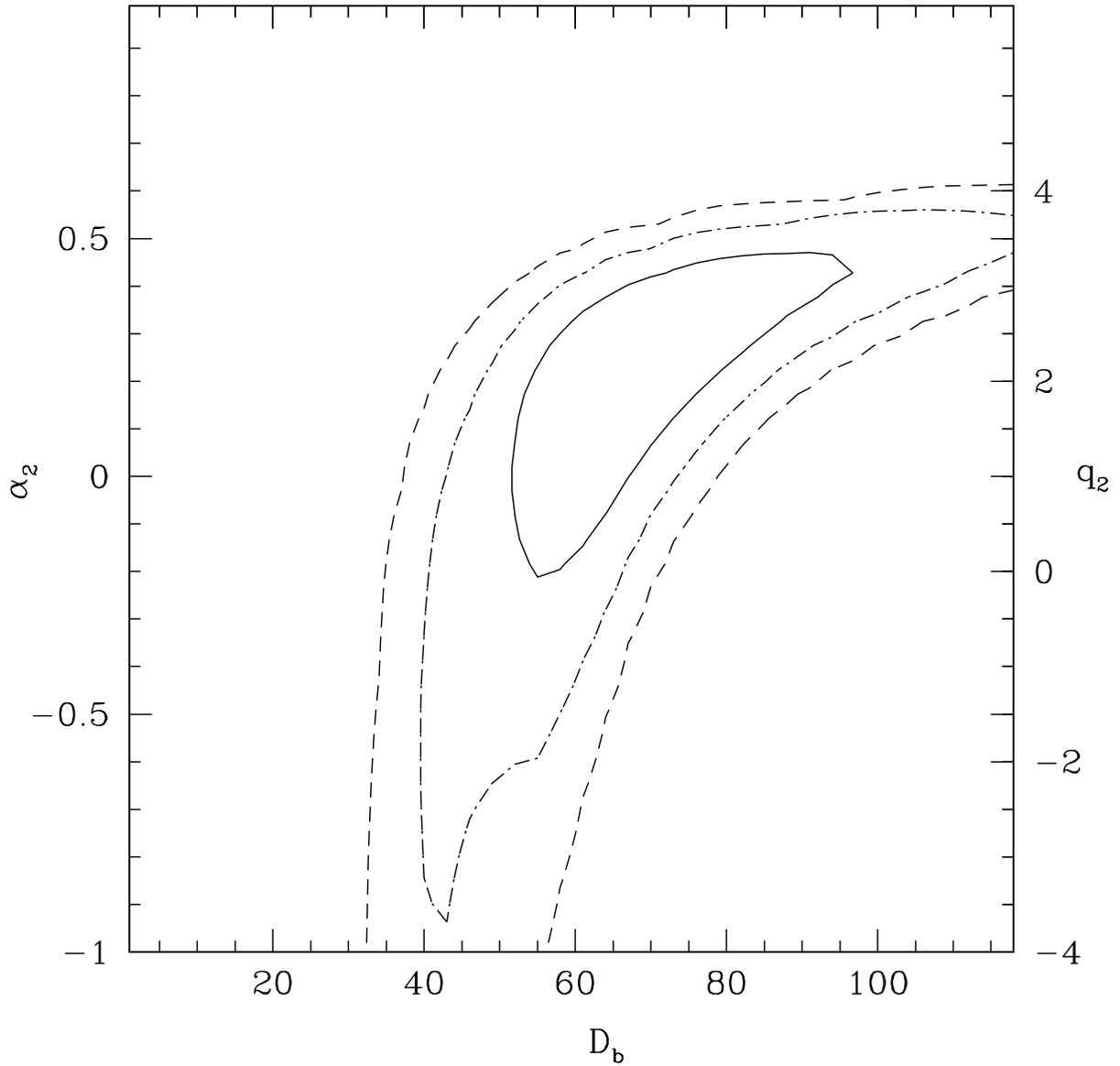} 
   \figcaption{The 1,2, and 3-sigma credible regions of the best-fit of Eq~\ref{eq:Fraser2008} to the F08b sample with $\alpha_1$ and $A$ held at their best-fit values. NOTE: This figure assumes that KBOs have 6\% albedos (see Section 4.1). \label{fig:a2Db}}   
\end{figure}

Presented in Figures~\ref{fig:a1ap}, \ref{fig:a1N}, and \ref{fig:a2Db} are the credible regions of the fits to the F08b sample. The function given by Equation~\ref{eq:Fraser2008} is derived from a size-distribution similar to that expected from accretion calculations, and thus provides a proper interpretation of the under-lying size-distribution. We find that the LF is consistent with a broken power-law size-distribution with large and small object slopes $q_1=4.8\pm3$ and $q_2=2\pm2$, and a break diameter, $D_b=62\pm 40$ km assuming 6\% albedos.

\begin{figure}[t] 
   \centering
   \plotone{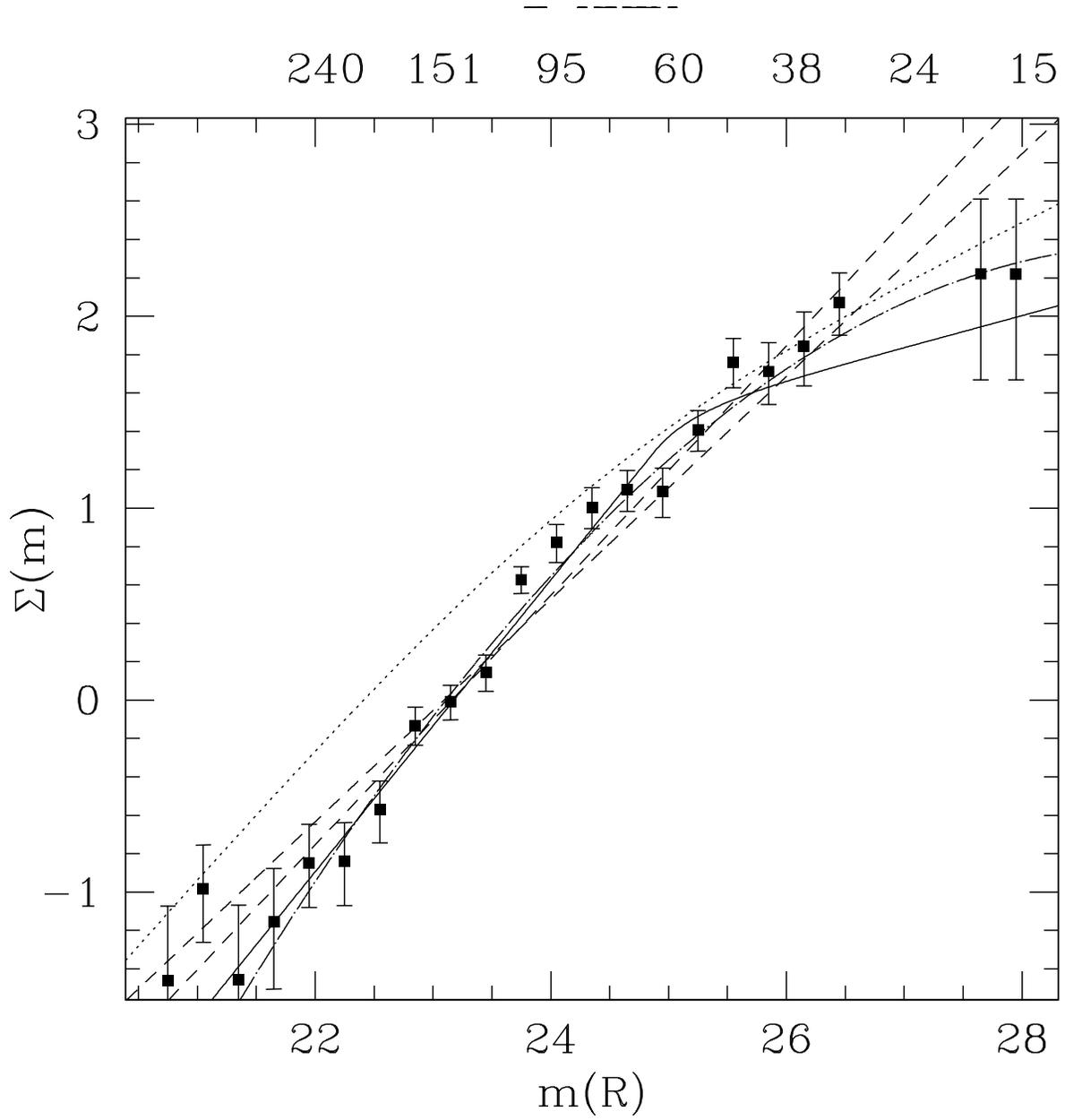} 
   \figcaption{Histogram of the data considered in the F08b sample with al magnitudes shifted to R-band using the colours presented in \citet{Fraser2008}. Errorbars are 1-sigma poisson intervals. The solid curve is the best-fit LF of Equation~\ref{eq:Fraser2008}. The dashed curves are the best-fit power-law LFs from this manuscript (shallow) and \citet{Fraser2008} (steep). The dashed-dotted curve is the best-fit rolling power-law. The dotted curve is the best-fit LF presented by \citet{Fuentes2008}. Object diameters (km) are given assuming 6\% albedos at 35 AU. \label{fig:diff}}   
\end{figure}

Presented in Figure~\ref{fig:diff} is a histogram of the F08b sample, the best-fit LFs for the three functional forms we consider, the best-fit power-law LF from \citet{Fraser2008}, and the best-fit LF presented by \citet{Fuentes2008}. As can be seen, power-laws do not provide acceptable fits to the data with $m(R)\sim28$, and a more complicated function is required. Equations~\ref{eq:rolling} and \ref{eq:Fraser2008} both provide reasonable descriptions of the observations.

The user is cautioned from drawing any further conclusions from Figure~\ref{fig:diff}. The F08b sample contains data from many different surveys. Thus the sample includes any calibration errors and sky-density variations between surveys which might lead the reader to a false conclusion about the structure of the LF (see \citet{Fraser2008} for a discussion of the magnitude and significance of these effects).

The best-fit LF presented by \citet{Fuentes2008} is a harmonic mean of two power-laws with bright-end slope $\alpha_1=0.7$ and faint-end slope $\alpha_2=0.3$ with a break magnitude of $R_{eq}\sim 24.3$. These parameters are consistent with the slopes and break magnitude of our best-fit non-power-law LFs. They find however, a factor of 3 higher sky-density which is inconsistent at greater than the 3-sigma level with the range of sky densities we find for similar bright-end slopes (see Figure~\ref{fig:a1N}). The source of the increased number of detections compared to that expected from fits to the F08b sample is unknown.  

\subsection{F08b$_{i>5}$ Sample \label{sec:highi_fit}}

When we consider the F08b$_{i>5}$ sample, we find that the best-fit power-law has slope $\alpha=0.55$ with $P(\Delta>\Delta_{obs})<0.04$, and is a poor description of the data. The sample has no objects fainter than $m(R)\sim 25.8$. The lack of faint objects requires a break near this point. The best-fit rolling power-law with parameters $(\alpha,\alpha',\Sigma_{23}) = (0.74,-0.06,0.46)$ is an acceptable fit to the data with $P(\Delta>\Delta_{obs})=0.1$. The best-fit broken power-law model is an acceptable description of the data with $P(\Delta>\Delta_{OBS})=0.1$. The log-likelihood ratio implies that the broken power-law model of Equation~\ref{eq:Fraser2008} is preferred over the rolling power-law at the 90\% significance level. The best-fit has $\alpha_1=0.7$ and breaks at $D_b=36$ km. Because the F08b$_{i>5}$ sample has no detections faint-ward of the break, both the break diameter and small-object slope of this sample are poorly constrained. The maximum likelihood is found with a break at $m(R)=25.8$ with no objects faint-ward of this point. ie. $\alpha_2=-\infty$. Note: we consider slopes, $\alpha_2>-6$ to avoid numerical errors in the likelihood calculations. 

\begin{figure}[t] 
   \centering
   \plotone{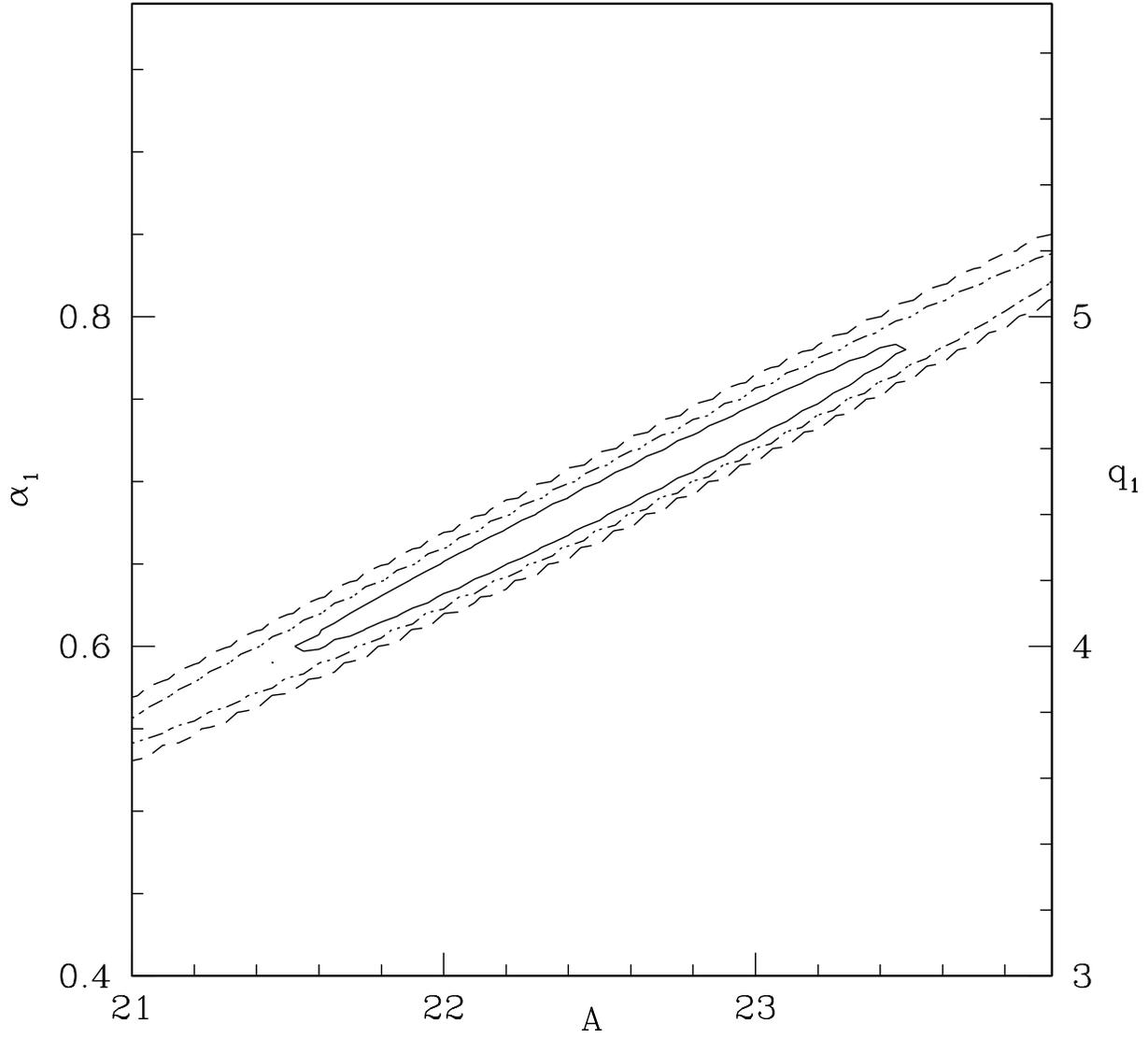} 
   \figcaption{The credible regions of the best-fit of Equation~\ref{eq:Fraser2008} to the F08b$_{i>5}$ sample. Contours are the 1, 2, and 3-sigma credible regions of the fit with $\alpha_2$ and $D_b$ held at their best-fit values.. \label{fig:a1N_highi}}   
\end{figure}

\begin{figure}[t] 
   \centering
   \plotone{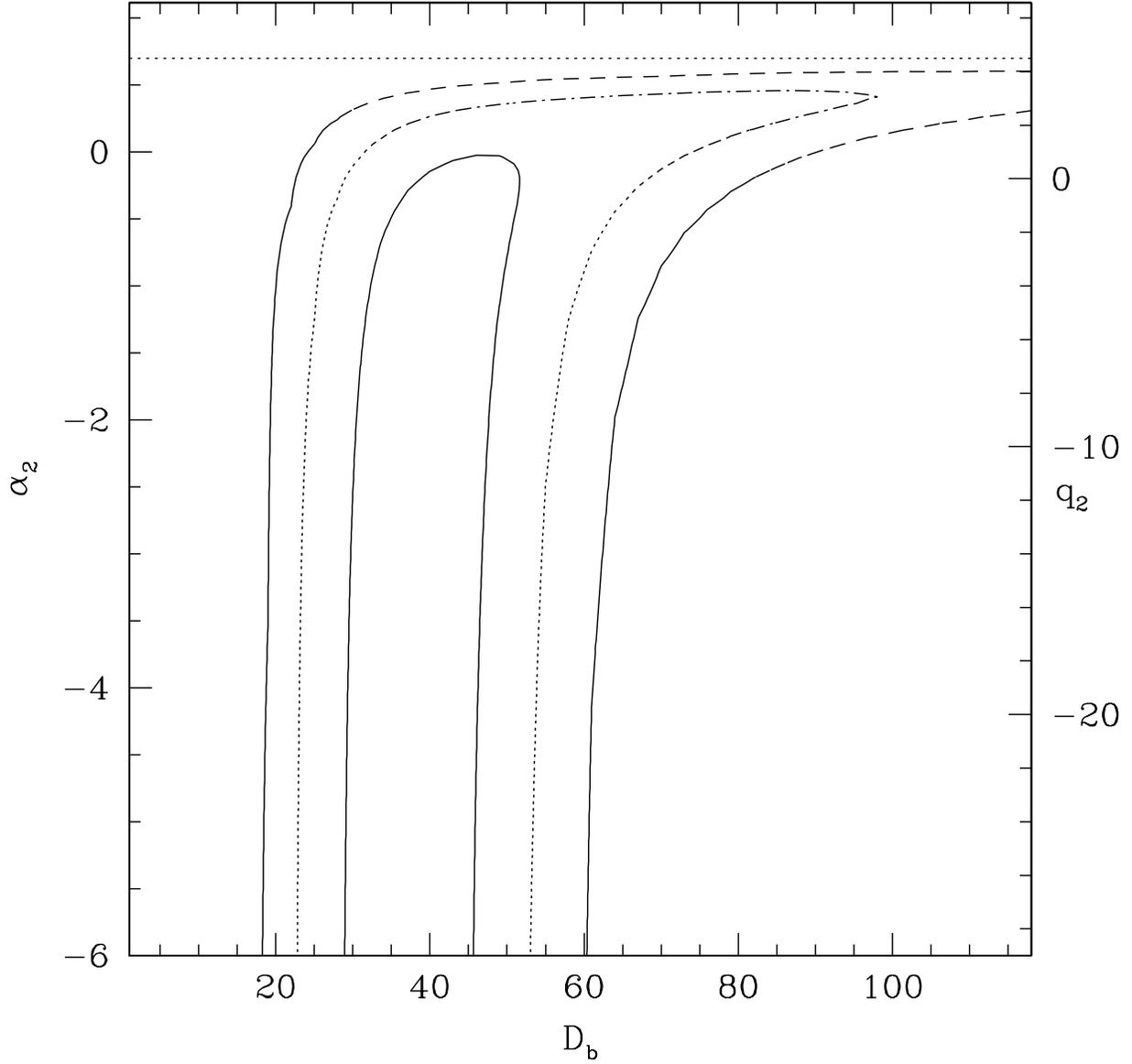} 
   \figcaption{The credible regions of the best-fit of Equation~\ref{eq:Fraser2008} to the F08b$_{i>5}$ sample. Contours are the 1, 2, and 3-sigma credible regions of the fit with $\alpha_1$ and $A$ held at their best-fit values. The dotted horizontal line marks the best-fit large object slope, and demonstrates the need for a break to shallower slopes for this sample.  NOTE: This figure assumes that KBOs have 6\% albedos (see Section 4.1). \label{fig:a2Db_highi}}   
\end{figure}

\begin{figure}[t] 
   \centering
   \plotone{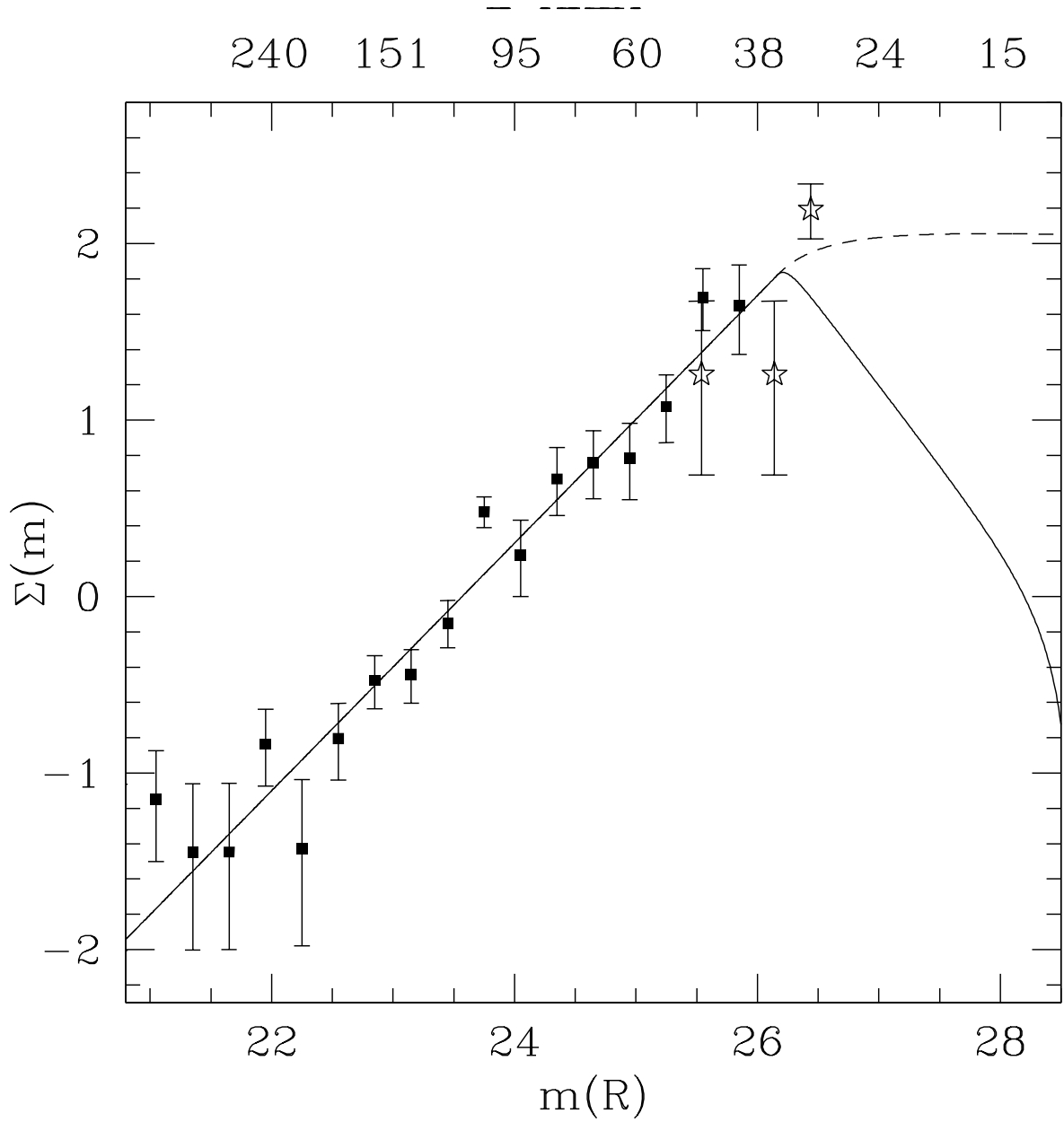} 
   \figcaption{Histogram of the data considered in the F08b$_{i>5}$ sample. Errorbars are 1-sigma poisson intervals. The solid curve is the best-fit of Equation~\ref{eq:Fraser2008}. The dashed curve is the LF expected when $\alpha_1$, $D_b$ and $A$ are held to their best-fit values, and $\alpha_2$ is taken at the upper 1-sigma limit, and represents the 1-sigma upper limit to the LF set by the absence of detections with $i>5$ from \citet{Bernstein2004}. The stars are the histogram of the Subaru survey presented in this manuscript. These points however, are unreliable as the inclinations for the objects discovered in this survey are highly uncertain. Object diameters (km) are given assuming 6\% albedos at 35 AU. \label{fig:diff_highi}}   
\end{figure}

Presented in Figures~\ref{fig:a1N_highi}, \ref{fig:a2Db_highi}, and \ref{fig:diff_highi} are the likelihood contours of the best-fit of Equation~\ref{eq:Fraser2008} to the F08b$_{i>5}$ sample and the histogram of those data. As can be seen, the LF is well described by a power-law for $m(R)\lesssim 25.8$. The lack of faint objects requires that a break must exist at $D_b\lesssim 100$ km. 

The faint-object slope is highly uncertain, with a 1-sigma upper-limit of $\alpha_2\sim0.6$. This fit however is consistent with that of the F08b sample best-fit at the $\sim1$-sigma level. Though the inclinations of the detections we present in this survey are inaccurate, they hint that the break in the LF for the F08b$_{i>5}$ sample is not as sharp as the fit would suggest. The detections from this survey are consistent with the same break diameter (36 km) that breaks to a much flatter faint object slope similar to the 1-sigma upper-limit of the fit $(\alpha_2\sim0)$. The fit to the F08b$_{i>5}$ sample however, constrains the break to be brightward of $m(R)\lesssim 26.8$. Additional data brightward of $m(R)\sim27.5$ with accurate inclinations is required before the break location can be accurately constrained. 

\subsection{F08b$_{i<5}$ Sample}
The low-inclination LF is sufficiently described by a power-law. The best-fit power-law is an acceptable description of the sample, with parameters $(\alpha,m_o)=(0.59,24.0)$ and an Anderson-Darling statistic, $P(\Delta>\Delta_{obs})=0.15$. Indeed the best-fit of Equation~\ref{eq:Fraser2008} is found when the break occurs at the bright-end limit of the data, suggesting there is no strong evidence for a break in the F08b$_{i<5}$ data.

\begin{figure}[t] 
   \centering
   \plotone{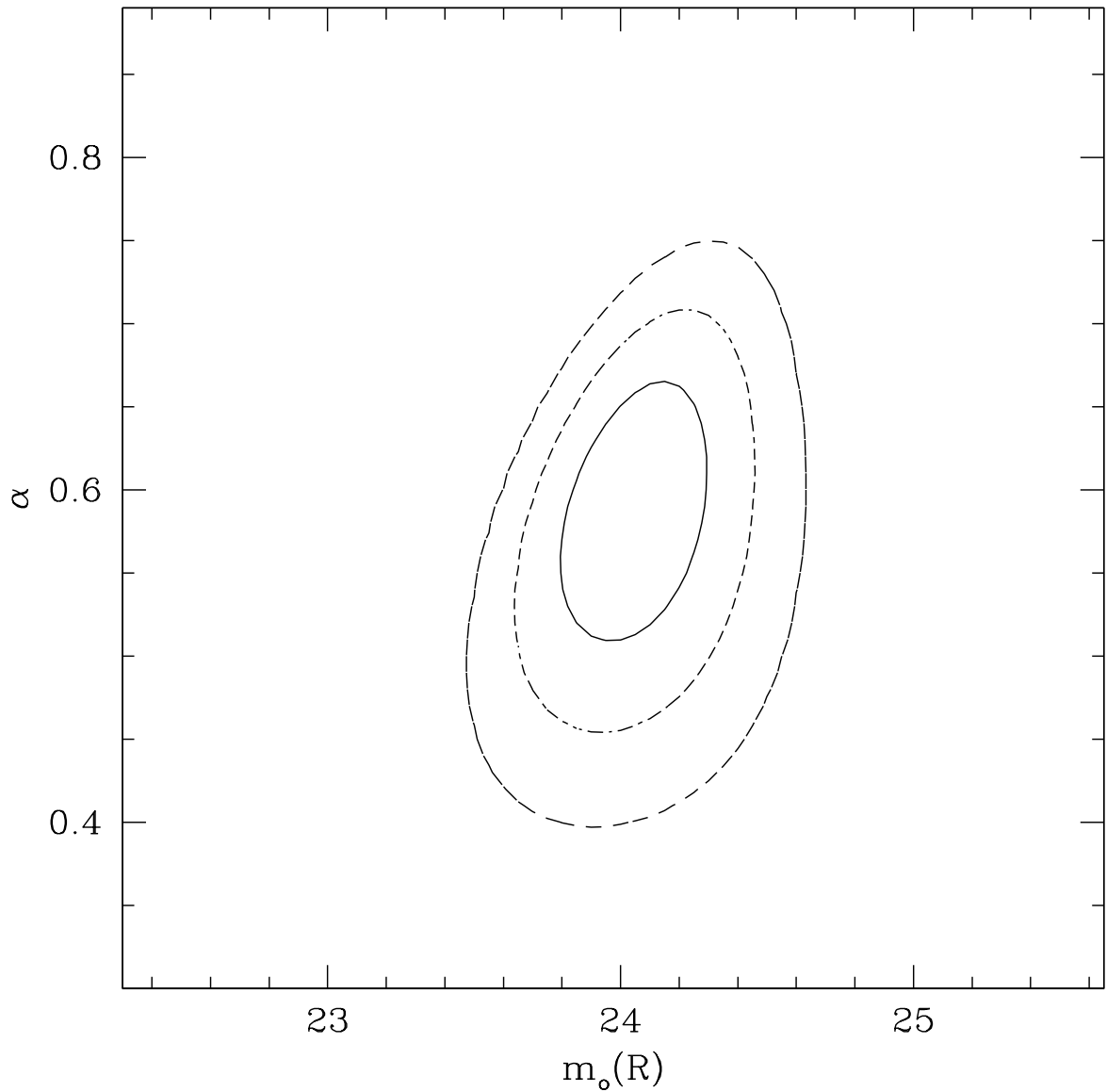} 
   \figcaption{The credible regions of the best-fit power-law to the F08b$_{i<5}$ sample. Contours are the 1, 2, and 3-sigma credible regions of the fit. The best-fit power-law is consistent with the collisional equilibrium slope $q=3.5$ and is a statistically sufficient description of this data-set. The best-fit broken power-law of the F08b$_{i>5}$ sample is a sufficient description of this data-set. \label{fig:amo_lowi}}   
\end{figure}

\begin{figure}[t] 
   \centering
   \plotone{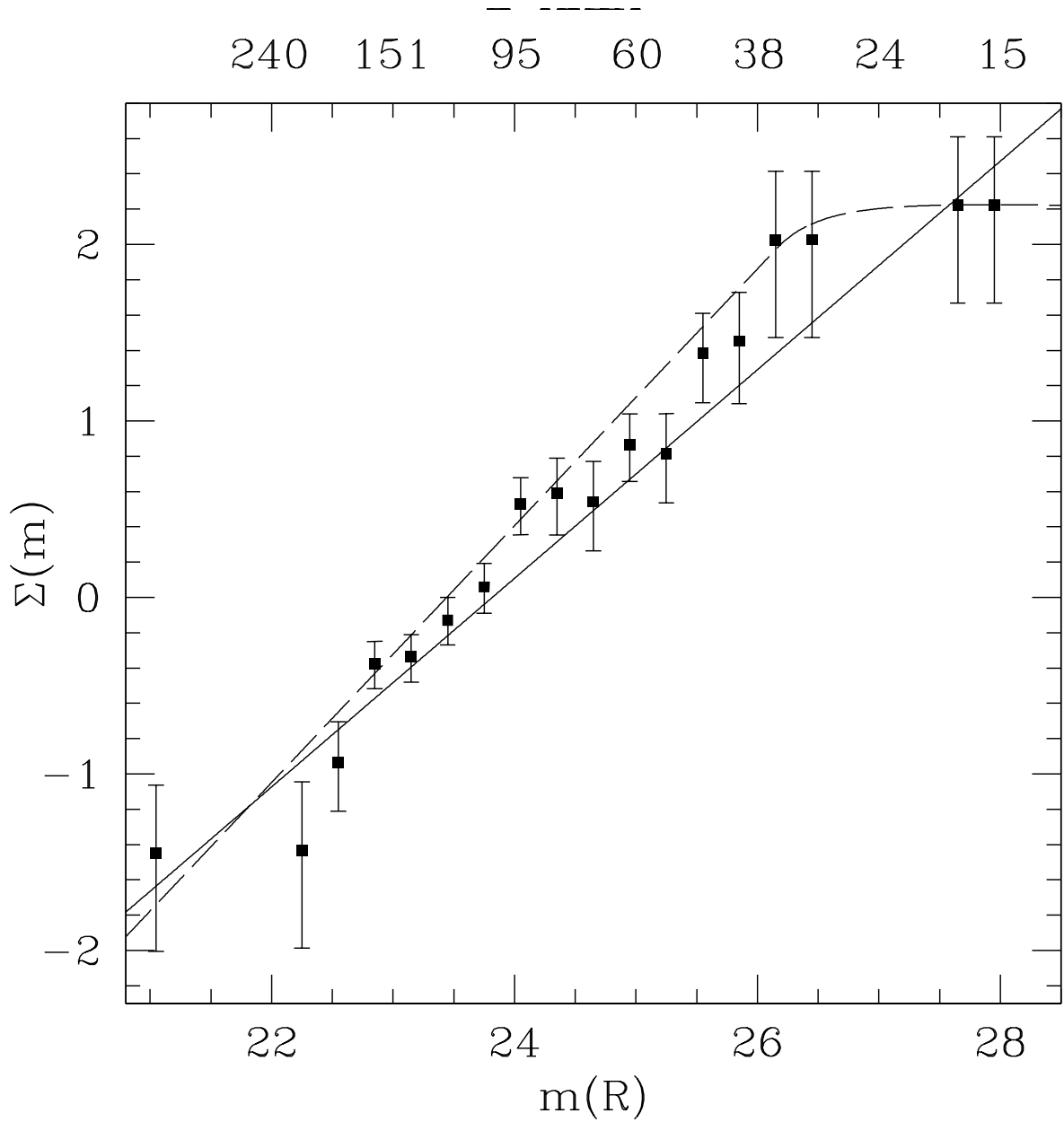} 
   \figcaption{Histogram of the data considered in the F08b$_{i<5}$ sample. Errorbars are 1-sigma poisson intervals. The solid curve is the best-fit power-law and is an adequate description of the $F08b_{i<5}$ sample. Object diameters (km) are given assuming 6\% albedos at 35 AU. The dashed line is a broken LF from Equation~\ref{eq:Fraser2008} with $(\alpha_1,\alpha_2,m_b,\log A) = (0.7,0,26.2,22.6)$ and is an adequate description of the data. The dashed curve is consistent at the 1-sigma level with the best-fit LF of the F08b$_{i>5}$ sample, demonstrating that, statistically speaking, the $F08b_{i<5}$ and $F08b_{i>5}$ samples have the same luminosity functions. \label{fig:diff_lowi}}   
\end{figure}

Presented in Figures~\ref{fig:amo_lowi} and \ref{fig:diff_lowi} are the likelihood contours of the fit to the low-inclination LF, and the histogram of the observations. As can be seen, there is no apparent evidence for a break in the magnitude range of the observations. 

To test whether or not the excited and cold samples have different LFs, we utilized the Anderson-Darling test described above (see Section~\ref{sec:analysis}) to determine whether or not the broken LF fit of the F08b$_{i>5}$ sample is an adequate description of the F08b$_{i<5}$ sample. We however, set $\alpha_2=0$ rather than the best-fit value, as this is a more realistic faint object slope than that of the best-fit of the F08b$_{i>5}$ sample $(\alpha_2\sim-\infty)$, as suggested by our survey. This slope is still within the 1-sigma credible region of the fit (see Section~\ref{sec:highi_fit}). We found that for the F08b$_{i<5}$ sample, an LF with $(\alpha_1,\alpha_2,m_b,\log A) =(0.7,0,26.2,22.7)$ has $P(\Delta>\Delta_{obs})=0.4$, and is a sufficient description of the  cold sample LF. We therefore conclude that we have no evidence that those KBOs with $i>5$ have a different LF than those with $i<5$.

\citet{Bernstein2004} found that the LFs of the cold and excited populations were different. \citet{Fuentes2008} found similar results for their survey alone, and concluded that the low inclination population LF exhibits a bright-end slope $\alpha\sim0.8-1.5$, steeper than $\alpha_1\sim0.7$ found when they considered objects of all $i$ in their survey. When they considered all available data - the F08b sample, but not including the survey presented here, and including their own search - they found that the low-inclination group exhibited a significantly steeper bright-object slope than the high-inclination sample. This finding is intriguing, as we find no evidence for a difference in the SDs of the low and high inclination groups. It is possible that their results are a consequence of not considering calibration and density offsets when performing the likelihood calculations, as we have done here. Clearly however, more observations are needed which probe the entire range of current observations before these results can be clarified.

\section{Discussion}
The best-fit parameters for the LF presented in Equation~\ref{eq:Fraser2008} imply that the size distribution is a power-law with slope $q_1\sim 4.8$ for large objects, which breaks to a shallower slope at diameter $D_b\sim 50-95$ km assuming 6\% albedos. Comparison of this size distribution to models which evolve a population of planetesimals and track their size distribution as a function of time in the Kuiper belt region can place a constraint on the formation processes and the duration of accretion in that region \citep{Kenyon2001, Kenyon2002, Kenyon2004}.

In the early stages of formation, run-away growth occurs, and very rapidly grow objects as large as $\sim 10^3$ km in the Kuiper belt \citep{Kenyon2002}. During this process, a steep-sloped large object size distribution develops, which flattens with time as more objects become ``large''. Calculations presented in \citet{Kenyon2002}, which simulate planet accretion for conditions in the Kuiper belt expected for the early solar system, imply that, in the absence of influences from Neptune, the modern-day large object slope would be shallower than that observed if accretion lasted the age of the solar system. They find that for an initial Kuiper belt mass similar to that predicted from the minimum-mass solar nebula model \citep{Hayashi1981} - much more massive than the current belt \citep{Fuentes2008} - accretion must have been halted after $\sim 100$ Myr. If KBOs are weak (easier to disrupt), then accretion might have gone on for as long as 1 Gyr \citep[see Figures 9, and 10 from][]{Kenyon2002}. Some event(s) must have halted this process by clearing the majority of initial mass out of the belt before the slope became too shallow.

\citet{Kenyon2004} has demonstrated that weaker bodies will exhibit a size distribution with a larger break diameter $D_b$ than stronger bodies would if they underwent the same evolutionary history. They calculated the size distribution expected from a Kuiper belt under the gravitational influence of Neptune which evolved for the age of the solar system. In this model, break diameters as large as 60 km were produced only for the weakest bodies they considered. This model however, produced a large object slope much too shallow to be consistent with the observations, implying that accretion over the age of the solar system cannot have occurred.

The existence of such a large break diameter in the KBO SD implies that, KBOs must be quite weak (strengthless rubble piles), and have undergone a period of increased collisional evolution. This suggests a scenario in which the event responsible for the early end to accretion, and the clearing of the majority of the mass in the Kuiper belt, also increased the rate of collisional evolution for a time, pushing the break diameter to the large value we see today.

Collisional disruption would be substantially increased for a period of time during the scattering of planetesimals by a rogue planet \citep{Gladman2006}, or during a close stellar passage \citep{Ida2000,Levison2004}. During these scenarios, relative velocities of KBOs are increased, causing collisions between bodies to result in catastrophic disruption, rather than accretion as would otherwise be the case. The combination of scattering and collisional grinding would produce a substantial mass loss in the Kuiper belt region, and produce the large knee observed currently. Increased bombardment would also occur if Neptune migrated outwards. In such a scenario, Neptune formed closer to the Sun than its current location. Gravitational scattering of small planetesimals transfer angular momentum to Neptune, causing the gas-giant to migrate outwards. During this process, some planetesimals have their orbits excited, and are thrust into the modern day Kuiper belt creating the orbital distribution we see today. The remaining planetesimals are cleared from the system during close encounters with Neptune causing a rapid and substantial mass-loss during the migration process. Various incarnations of this attractive scenario can provide the necessary collisional bombardment of KBOs, and simultaneously account for some of the other observed features of the Kuiper belt \citep{Malhotra1993,Levison2003,Hahn2005,Levison2008}.

The Neptune migration scenario predicts that the more excited population will have a more evolved size distribution, with a shallower large-object slope -  objects which originated closer to the Sun have stronger encounters with Neptune, and thus have the most excited final orbits. These  now excited objects would have originated from a more dense region, and hence underwent more rapid accretionary evolution, than objects which originated further from the Sun producing the shallower large object slope. We do not see any differences in the large object slopes of the cold and excited populations. Our findings suggest that, assuming migration of Neptune occurred, the total distance travelled by Neptune was sufficiently small such that the rate of planet formation and hence the size distribution of all objects scattered by Neptune and implaced in the Kuiper belt could not be substantially different.

\section{Conclusions}
We have performed a survey of the Kuiper covering $\sim 1/3$ a square degree of the sky using Suprime-cam on the Subaru telescope, to a limiting magnitude of $m_{50}(R)\sim 26.8$ and have found 36 new KBOs. Using the likelihood technique of \citet{Fraser2008} which accounts for calibration errors and sky density variations between separate observations, we have combined the observations of this survey with previous observations, and have found that the luminosity function of the Kuiper belt is inconsistent with a power-law with slope $\alpha_1=0.75$, but must break to a shallower slope $\alpha_2\sim0.2$ at magnitudes $m(R)\sim 24.1-25.3$. The luminosity function is consistent with a size distribution with large object slope $q_1\sim4.8$ that breaks to a shallower slope $q_2\sim1.9$ at a diameter of $\sim 50-95$ km assuming 6\% albedos. We have found no conclusive evidence that the size distribution of KBOs with $i<5$ is different than that of those with $i>5$.

\section{Acknowledgements}
This project was funded by the National Science and Engineering Research Council and the National Research Council of Canada. This research used the facilities of the Canadian Astronomy Data Centre operated by the National Research Council of Canada with the support of the Canadian Space Agency. This article is based in part on data collected at Subaru Telescope, which is operated by the National Astronomical Observatory of Japan.

\bibliographystyle{apj}
\bibliography{AstroElsart_P32008}

\begin{thebibliography}{37}
\expandafter\ifx\csname natexlab\endcsname\relax\def\natexlab#1{#1}\fi

\bibitem[{{Allen} {et~al.}(2002){Allen}, {Bernstein}, \&
  {Malhotra}}]{Allen2002}
{Allen}, R.~L., {Bernstein}, G.~M., \& {Malhotra}, R. 2002, AJ, 124, 2949

\bibitem[{{Astier} {et~al.}(2006){Astier}, {Guy}, {Regnault}, {Pain},
  {Aubourg}, {Balam}, {Basa}, {Carlberg}, {Fabbro}, {Fouchez}, {Hook},
  {Howell}, {Lafoux}, {Neill}, {Palanque-Delabrouille}, {Perrett}, {Pritchet},
  {Rich}, {Sullivan}, {Taillet}, {Aldering}, {Antilogus}, {Arsenijevic},
  {Balland}, {Baumont}, {Bronder}, {Courtois}, {Ellis}, {Filiol}, {Gon{\c
  c}alves}, {Goobar}, {Guide}, {Hardin}, {Lusset}, {Lidman}, {McMahon},
  {Mouchet}, {Mourao}, {Perlmutter}, {Ripoche}, {Tao}, \&
  {Walton}}]{Astier2006}
{Astier}, P., {Guy}, J., {Regnault}, N., {Pain}, R., {Aubourg}, E., {Balam},
  D., {Basa}, S., {Carlberg}, R.~G., {Fabbro}, S., {Fouchez}, D., {Hook},
  I.~M., {Howell}, D.~A., {Lafoux}, H., {Neill}, J.~D.,
  {Palanque-Delabrouille}, N., {Perrett}, K., {Pritchet}, C.~J., {Rich}, J.,
  {Sullivan}, M., {Taillet}, R., {Aldering}, G., {Antilogus}, P.,
  {Arsenijevic}, V., {Balland}, C., {Baumont}, S., {Bronder}, J., {Courtois},
  H., {Ellis}, R.~S., {Filiol}, M., {Gon{\c c}alves}, A.~C., {Goobar}, A.,
  {Guide}, D., {Hardin}, D., {Lusset}, V., {Lidman}, C., {McMahon}, R.,
  {Mouchet}, M., {Mourao}, A., {Perlmutter}, S., {Ripoche}, P., {Tao}, C., \&
  {Walton}, N. 2006, A\&A, 447, 31

\bibitem[{{Bernstein} \& {Khushalani}(2000)}]{Bernstein2000}
{Bernstein}, G. \& {Khushalani}, B. 2000, AJ, 120, 3323

\bibitem[{{Bernstein} {et~al.}(2004){Bernstein}, {Trilling}, {Allen}, {Brown},
  {Holman}, \& {Malhotra}}]{Bernstein2004}
{Bernstein}, G.~M., {Trilling}, D.~E., {Allen}, R.~L., {Brown}, M.~E.,
  {Holman}, M., \& {Malhotra}, R. 2004, AJ, 128, 1364

\bibitem[{{Bernstein} {et~al.}(2006){Bernstein}, {Trilling}, {Allen}, {Brown},
  {Holman}, \& {Malhotra}}]{Bernstein2005}
---. 2006, \aj, 131, 2364

\bibitem[{{Bertin} {et~al.}(2002){Bertin}, {Mellier}, {Radovich}, {Missonnier},
  {Didelon}, \& {Morin}}]{Bertin2002}
{Bertin}, E., {Mellier}, Y., {Radovich}, M., {Missonnier}, G., {Didelon}, P.,
  \& {Morin}, B. 2002, in Astronomical Society of the Pacific Conference
  Series, Vol. 281, Astronomical Data Analysis Software and Systems XI, ed.
  D.~A. {Bohlender}, D.~{Durand}, \& T.~H. {Handley}, 228--+

\bibitem[{{Brown} {et~al.}(2006){Brown}, {Schaller}, {Roe}, {Rabinowitz}, \&
  {Trujillo}}]{Brown2006}
{Brown}, M.~E., {Schaller}, E.~L., {Roe}, H.~G., {Rabinowitz}, D.~L., \&
  {Trujillo}, C.~A. 2006, ApJ, 643, L61

\bibitem[{{Fraser} \& {Kavelaars}(2008)}]{Fraser2008b}
{Fraser}, W.~C. \& {Kavelaars}, J. 2008, Submitted to Icarus

\bibitem[{{Fraser} {et~al.}(2008){Fraser}, {Kavelaars}, {Holman}, {Pritchet},
  {Gladman}, {Grav}, {Jones}, {Macwilliams}, \& {Petit}}]{Fraser2008}
{Fraser}, W.~C., {Kavelaars}, J.~J., {Holman}, M.~J., {Pritchet}, C.~J.,
  {Gladman}, B.~J., {Grav}, T., {Jones}, R.~L., {Macwilliams}, J., \& {Petit},
  J.-M. 2008, Icarus, 195, 827

\bibitem[{{Fuentes} \& {Holman}(2008)}]{Fuentes2008}
{Fuentes}, C.~I. \& {Holman}, M.~J. 2008, ArXiv e-prints, 804

\bibitem[{{Gladman} \& {Chan}(2006)}]{Gladman2006}
{Gladman}, B. \& {Chan}, C. 2006, \apjl, 643, L135

\bibitem[{{Gladman} {et~al.}(1998){Gladman}, {Kavelaars}, {Nicholson},
  {Loredo}, \& {Burns}}]{Gladman1998}
{Gladman}, B., {Kavelaars}, J.~J., {Nicholson}, P.~D., {Loredo}, T.~J., \&
  {Burns}, J.~A. 1998, AJ, 116, 2042

\bibitem[{{Gladman} {et~al.}(2001){Gladman}, {Kavelaars}, {Petit},
  {Morbidelli}, {Holman}, \& {Loredo}}]{Gladman2001}
{Gladman}, B., {Kavelaars}, J.~J., {Petit}, J.-M., {Morbidelli}, A., {Holman},
  M.~J., \& {Loredo}, T. 2001, AJ, 122, 1051

\bibitem[{{Gladman} {et~al.}(2008){Gladman}, {Marsden}, \&
  {Vanlaerhoven}}]{Gladman2008}
{Gladman}, B., {Marsden}, B.~G., \& {Vanlaerhoven}, C. 2008, {Nomenclature in
  the Outer Solar System} (The Solar System Beyond Neptune), 43--57

\bibitem[{{Gwyn}(2008)}]{Gwyn2008}
{Gwyn}, S.~D.~J. 2008, \pasp, 120, 212

\bibitem[{{Hahn} \& {Malhotra}(2005)}]{Hahn2005}
{Hahn}, J.~M. \& {Malhotra}, R. 2005, \aj, 130, 2392

\bibitem[{{Hayashi}(1981)}]{Hayashi1981}
{Hayashi}, C. 1981, Progress of Theoretical Physics Supplement, 70, 35

\bibitem[{{Ida} {et~al.}(2000){Ida}, {Larwood}, \& {Burkert}}]{Ida2000}
{Ida}, S., {Larwood}, J., \& {Burkert}, A. 2000, \apj, 528, 351

\bibitem[{{Jewitt} {et~al.}(1998){Jewitt}, {Luu}, \& {Trujillo}}]{Jewitt1998}
{Jewitt}, D., {Luu}, J., \& {Trujillo}, C. 1998, AJ, 115, 2125

\bibitem[{{Kenyon}(2002)}]{Kenyon2002}
{Kenyon}, S.~J. 2002, PASP, 114, 265

\bibitem[{{Kenyon} \& {Bromley}(2001)}]{Kenyon2001}
{Kenyon}, S.~J. \& {Bromley}, B.~C. 2001, AJ, 121, 538

\bibitem[{{Kenyon} \& {Bromley}(2004)}]{Kenyon2004}
---. 2004, AJ, 128, 1916

\bibitem[{{Kotz} \& {Johnson}(1983)}]{Kotz1983}
{Kotz}, S. \& {Johnson}, N. 1983, {Encyclopedia of Statistical Sciences} (John
  Wiley and Sons Ltd.)

\bibitem[{{Levison} \& {Morbidelli}(2003)}]{Levison2003}
{Levison}, H.~F. \& {Morbidelli}, A. 2003, \nat, 426, 419

\bibitem[{{Levison} {et~al.}(2004){Levison}, {Morbidelli}, \&
  {Dones}}]{Levison2004}
{Levison}, H.~F., {Morbidelli}, A., \& {Dones}, L. 2004, \aj, 128, 2553

\bibitem[{{Levison} {et~al.}(2008){Levison}, {Morbidelli}, {Vanlaerhoven},
  {Gomes}, \& {Tsiganis}}]{Levison2008}
{Levison}, H.~F., {Morbidelli}, A., {Vanlaerhoven}, C., {Gomes}, R., \&
  {Tsiganis}, K. 2008, Icarus, 196, 258

\bibitem[{{Malhotra}(1993)}]{Malhotra1993}
{Malhotra}, R. 1993, \nat, 365, 819

\bibitem[{{Miyazaki} {et~al.}(2002){Miyazaki}, {Komiyama}, {Sekiguchi},
  {Okamura}, {Doi}, {Furusawa}, {Hamabe}, {Imi}, {Kimura}, {Nakata}, {Okada},
  {Ouchi}, {Shimasaku}, {Yagi}, \& {Yasuda}}]{Miyazaki2002}
{Miyazaki}, S., {Komiyama}, Y., {Sekiguchi}, M., {Okamura}, S., {Doi}, M.,
  {Furusawa}, H., {Hamabe}, M., {Imi}, K., {Kimura}, M., {Nakata}, F., {Okada},
  N., {Ouchi}, M., {Shimasaku}, K., {Yagi}, M., \& {Yasuda}, N. 2002, \pasj,
  54, 833

\bibitem[{{Ouchi} {et~al.}(2004){Ouchi}, {Shimasaku}, {Okamura}, {Furusawa},
  {Kashikawa}, {Ota}, {Doi}, {Hamabe}, {Kimura}, {Komiyama}, {Miyazaki},
  {Miyazaki}, {Nakata}, {Sekiguchi}, {Yagi}, \& {Yasuda}}]{Ouchi2004}
{Ouchi}, M., {Shimasaku}, K., {Okamura}, S., {Furusawa}, H., {Kashikawa}, N.,
  {Ota}, K., {Doi}, M., {Hamabe}, M., {Kimura}, M., {Komiyama}, Y., {Miyazaki},
  M., {Miyazaki}, S., {Nakata}, F., {Sekiguchi}, M., {Yagi}, M., \& {Yasuda},
  N. 2004, \apj, 611, 660

\bibitem[{{Petit} {et~al.}(2006){Petit}, {Holman}, {Gladman}, {Kavelaars},
  {Scholl}, \& {Loredo}}]{Petit2006}
{Petit}, J.-M., {Holman}, M.~J., {Gladman}, B.~J., {Kavelaars}, J.~J.,
  {Scholl}, H., \& {Loredo}, T.~J. 2006, MNRAS, 365, 429

\bibitem[{{Pritchet}(2005)}]{Pritchet2005}
{Pritchet}, C.~J. 2005, in Astronomical Society of the Pacific Conference
  Series, Vol. 339, Observing Dark Energy, ed. S.~C. {Wolff} \& T.~R. {Lauer},
  60--+

\bibitem[{{Stansberry} {et~al.}(2007){Stansberry}, {Grundy}, {Brown},
  {Cruikshank}, {Spencer}, {Trilling}, \& {Margot}}]{Stansberry2007}
{Stansberry}, J., {Grundy}, W., {Brown}, M., {Cruikshank}, D., {Spencer}, J.,
  {Trilling}, D., \& {Margot}, J.-L. 2007, ArXiv Astrophysics e-prints

\bibitem[{{Stetson}(1989)}]{Stetson1989}
{Stetson}, P.~B. 1989, Advanced School of Astrophysics Universidade de Sao
  Paolo

\bibitem[{{Trujillo} \& {Brown}(2001)}]{Trujillo2001}
{Trujillo}, C.~A. \& {Brown}, M.~E. 2001, APJ, 554, L95

\bibitem[{{Trujillo} {et~al.}(2001){Trujillo}, {Jewitt}, \&
  {Luu}}]{Trujillo2001b}
{Trujillo}, C.~A., {Jewitt}, D.~C., \& {Luu}, J.~X. 2001, AJ, 122, 457

\bibitem[{{Yagi} {et~al.}(2002){Yagi}, {Kashikawa}, {Sekiguchi}, {Doi},
  {Yasuda}, {Shimasaku}, \& {Okamura}}]{Yagi2002}
{Yagi}, M., {Kashikawa}, N., {Sekiguchi}, M., {Doi}, M., {Yasuda}, N.,
  {Shimasaku}, K., \& {Okamura}, S. 2002, \aj, 123, 87

\bibitem[{{Yoshida} \& {Nakamura}(2007)}]{Yoshida2007}
{Yoshida}, F. \& {Nakamura}, T. 2007, \planss, 55, 1113

\end{thebibliography}

\begin{deluxetable}{lccccc}
   \tablecaption{Observation Details.\label{tab:observations}}
   \startdata	
      Night & Date & Seeing ('') & Number of Exposures & $\alpha$ & $\delta$ \\ \hline
      1 & April 22 2007 & $\sim0.4-0.7$ & 55 $\times$ 200 s. &  13:46:57.7 & -10:44:00\\
      2 & May 8 2007 & $\sim0.4-0.8$ & 53 $\times$ 200 s. &  13:46:57.7 & -10:44:00\\
       \enddata
\end{deluxetable}

\begin{deluxetable}{ccc}
   \tablecaption{Data Calibration Details.  The zero-point, $Z$, and colour term, $C$, of each chip of Suprime-Cam (see Section 2).  \label{tab:calibrations}}
   \startdata	
      Chip & $Z$ (mag.) & $C$ \\ \hline
      00 & 27.48 & -0.040 \\
      01 & 27.81 & -0.050 \\
      02 & 27.78 & -0.065 \\
      03 & 27.85 & -0.020 \\
      04 & 27.73 & -0.095 \\
      05 & 27.89 & -0.020 \\
      06 & 27.82 & -0.045 \\
      07 & 27.83 & -0.040 \\
      08 & 27.86 & -0.040 \\
      09 & 27.75 & -0.065 \\
   \enddata
\end{deluxetable}
   
\begin{deluxetable}{lccccc}
   \tablecaption{Survey Results. $r_{SDSS}$ is the average r-band magnitude, $n$ is the number of flux measurements used in the average, $\Delta$ is the object's heliocentric distance, and $i$ is the object's inclination. \label{tab:objects}}
   \startdata	
   Object & $ r_{SDSS}$ (mag.) & $n$ & $\Delta$ & $i$ ($^o$) & Follow-up \\ \hline
   c9a2 & $24.038\pm0.0424$ &  3 & $40\pm3$ & $22\pm14$ & n \\
   c1a3 & $24.368\pm0.0575$ &  3 & $44\pm3$ & $9\pm12$ & n \\
   c6a3 & $24.394\pm0.0585$ &  6 & $45\pm2$ & $5\pm2$ & y \\
   c1a1 & $24.991\pm0.1018$ &  3 & $44\pm3$ & $1\pm7$ & n \\
   c3a1 & $25.179\pm0.1212$ &  3 & $42\pm3$ & $0\pm2$ & n \\
   c5a2 & $25.245\pm0.1293$ &  3 & $41\pm3$ & $3.3\pm10$ & n \\
   c5a5 & $25.373\pm0.1444$ &  3 & $49\pm3$ & $3\pm14$ & n \\
   c8a2 & $25.397\pm0.1485$ &  3 & $47\pm4$ & $15\pm15$ & n \\
   c4a1 & $25.432\pm0.1527$ &  3 & $45\pm3$ & $9\pm13$ & n \\
   c9a1 & $25.434\pm0.1527$ &  3 & $35\pm3$ & $28\pm16$ & n \\
   c9a4 & $25.438\pm0.1541$ &  3 & $48\pm3$ & $2\pm13$ & n \\
   c8a1 & $25.458\pm0.1569$ &  3 & $43\pm3$ & $2\pm11$ & n \\
   c2a1 & $25.462\pm0.1569$ &  3 & $48\pm3$ & $7\pm14$ & n \\
   c4a3 & $25.569\pm0.1737$ &  3 & $38\pm3$ & $5\pm10$ & n \\
   c4a5 & $25.823\pm0.2187$ &  3 & $42\pm3$ & $5\pm11$ & n \\
   c5a9 & $25.875\pm0.2311$ &  3 & $45\pm3$ & $5\pm12$ & n \\
   c6a1 & $26.019\pm0.2629$ &  6 & $37\pm3$ & $2\pm9$ & y \\
   c0a2 & $26.068\pm0.2753$ &  6 & $45\pm2$ & $1.1\pm0.3$ & y \\
   c2a3 & $26.279\pm0.3340$ &  4 & $43\pm3$ & $5\pm12$ & y \\
   c1a4 & $26.282\pm0.3340$ &  3 & $44\pm3$ & $4\pm12$ & n \\
   c0a1 & $26.334\pm0.3498$ &  6 & $45\pm3$ & $14\pm5$ & y \\
   c6a2 & $26.382\pm0.3663$ &  6 & $44\pm2$ & $3\pm1$ & y \\
   c2a2 & $26.583\pm0.4404$ &  6 & $39\pm2$ & $10\pm4$ & y \\
   c4a6 & $26.610\pm0.4527$ &  3 & $43\pm4$ & $23\pm17$ & n \\
   c4a2 & $26.673\pm0.4784$ &  3 & $45\pm3$ & $12\pm13$ & n \\
   c1a5 & $26.680\pm0.4828$ &  2 & $37\pm3$ & $27\pm17$ & n \\
   c5a3 & $26.688\pm0.4873$ &  2 & $44\pm3$ & $7\pm12$ & n \\
   c5a8 & $26.737\pm0.5103$ &  3 & $47\pm3$ & $7\pm13$ & n \\
   c5a7 & $26.858\pm0.5699$ &  3 & $44\pm3$ & $1\pm9$ & n \\
   c4a7 & $26.859\pm0.5699$ &  3 & $34\pm3$ & $9\pm9$ & n \\
   c9a3 & $26.883\pm0.5805$ &  1 & $47\pm5$ & $75\pm29$ & n \\
   c7a1 & $26.925\pm0.6079$ &  3 & $45\pm3$ & $16\pm15$ & n \\
   c5a1 & $27.120\pm0.7241$ &  3 & $43\pm3$ & $1\pm8$ & n \\
   c5a6 & $27.276\pm0.8391$ &  3 & $47\pm3$ & $6\pm14$ & n \\
   c1a2 & $27.353\pm0.8950$ &  2 & $55\pm4$ & $12\pm18$ & n \\
   c5a4 & $28.807\pm3.4344$ &  3 & $42\pm3$ & $0\pm1$ & n \\
   \enddata
\end{deluxetable}

\footnotesize
\begin{landscape}
\begin{deluxetable}{lcccccc}
	\tablecaption{Luminosity Function Fits (LF and SD parameters) and associated Anderson-Darling statistics. $<$ indicate 1-sigma upper  limits. We do not present fits utilizing functions with more degrees of freedom than are statistically warranted by the observations. The F08b$_{i<5}$ sample was not fit with Equation~\ref{eq:Fraser2008} but is adequately described by LF parameters within the 1-sigma range of the fit of the F08b$_{i>5}$ sample. NOTE: $q_2=-\infty$ represents a complete absence of objects smaller than the break diameter. \label{tab:fits}}
	\tablehead{ \colhead{Data-set} & \multicolumn{2}{c}{Power-Law} & \multicolumn{2}{c}{Rolling Power-law} & \multicolumn{2}{c}{Broken Power-law}}

	\startdata
	 & $(\alpha,m_o)$ & $P(\Delta>\Delta_{obs})$ & $(\Sigma_{23}, \alpha, \alpha')$ & $P(\Delta>\Delta_{obs})$ & $(\log A,\alpha_1,\alpha_2,m_b)$ & $P(\Delta>\Delta_{obs})$\\  
	 &   & &   & & $(\log A, q_1, q_2, D_b)$ & \\\hline
		F08b & $(0.58,23.31)$ & $<0.04$ & $(0.82,0.80,-0.06)$ & 0.4 & $(23.6,0.76, 0.18, 24.9)$ & 0.4 \\
		& & & & & $(23.6,4.8, 1.9, 62)$ &  \\
		F08b$_{i<5}$ & $(0.59,24.0)$ & 0.15 & $-$ & - & - & - \\
		F08b$_{i>5}$ & $(0.55,23.81)$ & $<0.04$ & $(0.46,0.74,-0.06)$ & 0.1 & $(22.7,0.70,-6,26.2)$ & 0.1 \\
		& & & & & $(22.7,4.5,-\infty,36)$ &  \\
	\enddata
\end{deluxetable}

\end{landscape}
\normalsize

\end{document}